\documentclass[authoryear,5p,longtitle]{elsarticle}
\usepackage{amsmath}
\usepackage{amssymb}
\usepackage{amsthm}
\usepackage{graphicx}
\usepackage[colorinlistoftodos]{todonotes}
\usepackage[colorlinks=true, allcolors=blue]{hyperref}
\usepackage{natbib}
\usepackage{geometry}

\usepackage{lineno}
\newcommand{\snana}{{{\tt SNANA}}}

\newcommand{\as}{{\tt autoScan}}

\newcommand{\diffimg}{{\tt DiffImg}}

\usepackage{subfigure}
\usepackage{ulem}
\newcommand{\arcsec}{^{\prime\prime}}
\newcommand{\ldquo}{``}
\newcommand{\rdquo}{''}
\journal{Astronomy and Computing}

\begin{document}
\begin{frontmatter}
\title{Optical follow-up of gravitational wave triggers with DECam during the first two LIGO/VIRGO observing runs}

\author[fnal]{K.~Herner}
\ead{kherner@fnal.gov}
\author[fnal]{J.~Annis}
\author[upenn,3]{D.~Brout}
\author[brand,fnal]{M.~Soares-Santos}
\author[uofc]{R.~Kessler}
\author[upenn]{M.~Sako}
\author[fnal,uofc]{R.~Butler}
\author[uofc]{Z.~Doctor} 
\author[fnal,ucl]{A.~Palmese}
\author[fnal]{S.~Allam}
\author[fnal]{D.~L.~Tucker}
\author[linea,camp]{F.~Sobreira}
\author[fnal]{B.~Yanny}
\author[fnal]{H.~T.~Diehl}
\author[fnal,uofc]{J.~Frieman}
\author[fnal,scar]{N.~Glaeser}
\author[brand]{A.~Garcia}
\author[fnal,uiucp]{N.~F.~Sherman}
\author[lsst,wisc]{K.~Bechtol}
\author[cfa]{E.~Berger}
\author[cfa,uofc]{H.~Y.~Chen}
\author[nott]{C.~J.~Conselice}
\author[tamu]{E.~Cook}
\author[cfa]{P.~S.~Cowperthwaite}
\author[uq]{T.~M.~Davis}
\author[fnal]{A.~Drlica-Wagner}
\author[uofc]{B.~Farr}
\author[fnal]{D.~Finley}
\author[ucsc]{R.~J.~Foley}
\author[ift]{J.~Garcia-Bellido}
\author[kavli]{M.~S.~S.~Gill}
\author[uiuca,ncsa]{R.~A.~Gruendl}
\author[uofc]{D.~E.~Holz}
\author[fnal]{N.~Kuropatkin}
\author[fnal]{H.~Lin}
\author[fnal]{J.~Marriner}
\author[tamu]{J.~L.~Marshall}
\author[noia]{T.~Matheson}
\author[fnal]{E.~Neilsen}
\author[uiuca,ncsa]{F.~Paz-Chinch\'on}
\author[tamu]{M.~Sauseda}
\author[uofc]{D.~Scolnic}
\author[cfa]{P.~K.~G.~Williams}
\author[ift]{S.~Avila}
\author[cnrs,sorb]{E.~Bertin}
\author[fnal]{E.~Buckley-Geer}
\author[slac]{D.~L.~Burke}
\author[linea,ciemat]{A.~Carnero~Rosell}
\author[uiuca,ncsa]{M.~Carrasco-Kind}
\author[ifae]{J.~Carretero}
\author[linea,obnat]{L.~N.~da Costa}
\author[ciemat]{J.~De Vicente}
\author[hyder]{S.~Desai}
\author[ucl]{P.~Doel}
\author[caltech,jpl]{T.~F.~Eifler}
\author[ucsc]{S.~Everett}
\author[iss]{P.~Fosalba}
\author[iss]{E.~Gaztanaga}
\author[micha,michp]{D.~W.~Gerdes}
\author[linea,obnat]{J.~Gschwend}
\author[fnal]{G.~Gutierrez}
\author[ucl,eth]{W.~G.~Hartley}
\author[ucsc]{D.~L.~Hollowood}
\author[osua,osu]{K.~Honscheid}
\author[cfa]{D.~J.~James}
\author[kavli]{E.~Krause}
\author[aao]{K.~Kuehn}
\author[ucl]{O.~Lahav}
\author[fnal]{T.~S.~Li}
\author[linea,usp]{M.~Lima}
\author[linea,obnat]{M.~A.~G.~Maia}
\author[upenn]{M.~March}
\author[uiuca,ncsa]{F.~Menanteau}
\author[icrea,ifae]{R.~Miquel}
\author[prince]{A.~A.~Plazas}
\author[ciemat]{E.~Sanchez}
\author[fnal]{V.~Scarpine}
\author[michp]{M.~Schubnell}
\author[iss,ieec]{S.~Serrano}
\author[ciemat]{I.~Sevilla-Noarbe}
\author[southampton]{M.~Smith}
\author[ornl]{E.~Suchyta}
\author[michp]{G.~Tarle}
\author[fnal]{W.~Wester}
\author[fnal]{Y.~Zhang}

\address[fnal]{Fermi National Accelerator Laboratory, P. O. Box 500, Batavia, IL 60510, USA}
\address[upenn]{Department of Physics and Astronomy, University of Pennsylvania, Philadelphia, PA 19104, USA}
\address[3]{NASA Einstein Fellow}
\address[brand]{Department of Physics, Brandeis University, 415 South Street, Waltham, MA 02453, USA}
\address[uofc]{Kavli Institute for Cosmological Physics, University of Chicago, Chicago, IL 60637, USA}
\address[ucl]{Department of Physics \& Astronomy, University College London, Gower Street, London, WC1E 6BT, UK}
\address[linea]{Laboratorio Interinstitucional de e-Astronomia, Rua Gal. Jos\'{e} Cristino 77, Rio de Janeiro, RJ - 20921-400, Brazil}
\address[camp]{Instituto de F\'isica Gleb Wataghin, Universidade Estadual de Campinas, 13083-859, Campinas, SP, Brazil}
\address[scar]{University of South Carolina, Columbia, SC 29201, USA}
\address[uiucp]{Department of Physics, University of Illinois at Urbana-Champaign, 1110 W. Green Street, Urbana, IL 61801, USA}
\address[lsst]{LSST, 933 North Cherry Avenue, Tucson, AZ 85721, USA}
\address[wisc]{Physics Department, 2320 Chamberlin Hall, University of Wisconsin-Madison, 1150 University Avenue Madison, WI 53706-1390, USA}
\address[cfa]{Harvard-Smithsonian Center for Astrophysics, Cambridge, MA 02138, USA}
\address[nott]{Centre for Astronomy and Particle Theory, School of Physics \& Astronomy, University of Nottingham, Nottingham, NG7 2RD UK}
\address[tamu]{George P. and Cynthia Woods Mitchell Institute for Fundamental Physics and Astronomy, and Department of Physics and Astronomy, Texas A\&M University, College Station, TX 77843, USA}
\address[uq]{School of Mathematics and Physics, The University of Queensland, Brisbane, QLD 4072, Australia}
\address[ucsc]{Santa Cruz Institute for Particle Physics, Santa Cruz, CA 95064, USA}
\address[ift]{Instituto de Fisica Teorica UAM/CSIC, Universidad Autonoma de Madrid, Cantoblanco 28049 Madrid, Spain}
\address[kavli]{Kavli Institute for Particle Astrophysics \& Cosmology, P. O. Box 2450, Stanford University, Stanford, CA 94305, USA}
\address[uiuca]{Department of Astronomy, University of Illinois at Urbana-Champaign, 1002 W. Green Street, Urbana, IL 61801, USA}
\address[ncsa]{National Center for Supercomputing Applications, 1205 West Clark St., Urbana, IL 61801, USA}
\address[noia]{NSF’s National Optical Infrared Astronomy Research Laboratory, 950 North Cherry Avenue, Tucson, AZ 85719, USA}
\address[cnrs]{CNRS, UMR 7095, Institut d'Astrophysique de Paris, F-75014, Paris, France}
\address[sorb]{Sorbonne Universit\'{e}s, UPMC Univ Paris 06, UMR 7095, Institut d’Astrophysique de Paris, F-75014, Paris, France}
\address[slac]{SLAC National Accelerator Laboratory, Menlo Park, CA 94025, USA}
\address[ciemat]{Centro de Investigaciones Energ\'eticas, Medioambientales y Tecnol\'ogicas (CIEMAT), Madrid, Spain}
\address[ifae]{Institut de F\'isica d'Altes Energies (IFAE), The Barcelona Institute of Science and Technology, Campus UAB, 08193 Bellaterra (Barcelona), Spain}
\address[obnat]{Observat\'orio Nacional, Rua Gal. Jos\'e Cristino 77, Rio de Janeiro,
RJ - 20921-400, Brazil}
\address[hyder]{Department of Physics, IIT Hyderabad, Kandi, Telangana 502285, India}
\address[caltech]{Department of Physics, California Institute of Technology, Pasadena, CA 91125, USA}
\address[jpl]{Jet Propulsion Laboratory, California Institute of Technology, 4800 Oak Grove Drive, Pasadena, CA 91109, USA}
\address[iss]{Institute of Space Sciences, IEEC-CSIC, Campus UAB, Carrer de Can Magrans, s/n, 08193 Barcelona, Spain}
\address[micha]{Department of Astronomy, University of Michigan, Ann Arbor, MI 48109, USA}
\address[michp]{Department of Physics, University of Michigan, Ann Arbor,
MI 48109, USA}
\address[eth]{Department of Physics, ETH Z\"{u}rich, Wolfgang-Pauli-Strasse 16, CH-8093 Z\"{u}rich, Switzerland}
\address[osua]{Center for Cosmology and Astro-Particle Physics, The Ohio State University, Columbus, OH 43210, USA}
\address[osu]{Department of Physics, The Ohio State University, Columbus, OH 43210, USA}
\address[aao]{Australian Astronomical Observatory, North Ryde, NSW 2113, Australia}
\address[usp]{Departamento de F\'isica Matem\'atica, Instituto de F\'isica, Universidade de S\~ao Paulo, CP 66318, S\~ao Paulo, SP, 05314-970, Brazil}
\address[icrea]{Instituci\'o Catalana de Recerca i Estudis Avan\c{c}ats, E-08010 Barcelona, Spain}
\address[prince]{Department of Astrophysical Sciences, Princeton University, Peyton Hall, Princeton, NJ 08544, USA}
\address[ieec]{Institut d'Estudis Espacials de Catalunya (IEEC), 08034, Barcelona, Spain}
\address[southampton]{School of Physics and Astronomy, University of Southampton, Southampton, SO17 1BJ, UK}
\address[ornl]{Computer Science and Mathematics Division, Oak Ridge National Laboratory, Oak Ridge, TN 37831, USA}


\begin{abstract}
Gravitational wave (GW) events detectable by LIGO and Virgo have several possible progenitors, including black hole mergers, neutron star mergers, black hole--neutron star mergers, supernovae, and cosmic string cusps. A subset of GW events are expected to produce electromagnetic (EM) emission that, once detected, will provide complementary information about their astrophysical context. To that end, the LIGO--Virgo Collaboration (LVC) sends GW candidate alerts to the astronomical community so that searches for their EM counterparts can be pursued. The DESGW group, consisting of members of the Dark Energy Survey (DES), the LVC, and other members of the astronomical community, uses the Dark Energy Camera (DECam) to perform a search and discovery program for optical signatures of LVC GW events. DESGW aims to use a sample of GW events as standard sirens for cosmology. 
Due to the short decay timescale of the expected EM counterparts and the need to quickly eliminate survey areas with no counterpart candidates, it is critical to complete the initial analysis of each night's images as quickly as possible. 
We discuss our search area determination, imaging pipeline, and candidate selection processes. We review results from the DESGW program during the first two LIGO--Virgo observing campaigns and introduce other science applications that our pipeline enables.
\end{abstract}

\end{frontmatter}



\section{Introduction}

A particular challenge for precision cosmology is the absolute calibration of cosmic distances. 
Traditionally, it employs a cosmic distance ladder: direct geometrical parallax measurements of 
nearby stars that calibrate indirect measurements for larger distances, in a series of
steps extending to cosmological scales \citep{2019ApJ...876...85R}. This indirect approach introduces systematic uncertainties 
to measurements from type-Ia supernovae 
and other probes. Gravitational wave (GW) signals from merging binary objects, 
however, act as standard sirens, enabling distances to be measured directly to the source without relying on a cosmic distance ladder
\citep{1986Natur.323..310S, Holz2005, 2012PhRvD..86d3011D}. 
This fact motivates the pursuit of
a cosmology program using GW events. 
The results reported by the LIGO--Virgo Collaboration (LVC) in their first observing run with an advanced interferometer network included detection of binary black hole (BBH)  mergers at $440^{+180}_{-190}$
\citep{2016PhRvL.116x1103A} and  $420^{+160}_{-180}$ \citep{2016PhRvL.116f1102A}
Mpc, proving that the current generation of interferometers are sensitive enough to observe mergers at cosmological distances.

Anticipating that such sensitivity would be in reach, the Dark Energy Survey (DES) launched its GW program (DESGW) in the first LVC run (O1, 2015--2016). DESGW is a search and discovery program for electromagnetic (EM) signatures of GW events. While there are multiple reasons to search for GW electromagnetic counterparts, our primary goal is to perform a measurement of the Hubble Constant with the best precision afforded by the available data.
The second LVC run (O2, 2016--2017) saw the observation of a binary neutron star (BNS) merger \citep{2017PhRvL.119p1101A} and DESGW contributed \citep{2017ApJ...848L..16S} to the discovery of its EM counterpart \citep{2017ApJ...848L..12A}. The third run (O3, 2019--2020) is ongoing at the time of this writing.
The DESGW cosmology program has two different components: 1) If the GW signal has an EM counterpart, the location of the event may be identified to the arcsecond and a redshift for the galaxy obtained. 2) If the GW counterpart is electromagnetically dark, then
the redshift must be obtained statistically using all the galaxies in the probable localization volume of the event.

In this article we detail, for the first time, the DESGW observational program to search for EM counterparts of GW events in the optical range using the Dark Energy Camera (DECam). DESGW has been active during all three LVC observing runs to date and we document changes implemented from one run to another. We begin this article with a discussion of cosmology with standard sirens (Section~\ref{sec:sirens}), and then describe DECam itself (Section~\ref{sec:decam}), along with how the decision to trigger on a given GW event is made, including the interaction with other telescope users. Section~\ref{sec:exsetup} describes the experimental setup, and Section~\ref{sec:mapmaking} describes our overall observing strategy and construction of a detailed observing plan for each night. Sections~\ref{sec:proc} though \ref{sec:grid} discuss our image processing pipeline and final candidate selection. We conclude with a brief review of DESGW results from the first two observing seasons in Section~\ref{sec:results} and a discussion of planned improvements to our program and other science applications in Section~\ref{sec:future}.

\section{Cosmology with standard sirens}\label{sec:sirens}
\subsection{Bright Standard Sirens}

Mergers for which both gravitational and electromagnetic emission
are detectable, {\it bright standard sirens}, can be used
to determine cosmological parameters such as $H_0$ via the distance-redshift relation: the distance measurement comes from the GW signal and the redshift from identifying the host
galaxy of the EM counterpart.
Mergers of neutron stars, or of a neutron star and a black hole,
 have multiple predicted signatures
\citep{2010MNRAS.406.2650M,2010ApJ...725..496N,2011PhRvL.107j1102G,2014ARA&A..52...43B,2017LRR....20....3M,2017CQGra..34j4001R,2018ApJ...852..109T,2018MNRAS.473..576G,PhysRevD.100.043001,2019arXiv190504495D}:
a gravitational wave chirp,
a neutrino burst,
a gamma-ray burst followed by an afterglow in various wavelengths, and
an optical transient referred in the literature as a {\it kilonova}, which we aim to detect with DECam.
Fainter, redder, and shorter-lived than supernovae, kilonovae are
challenging from the observational point of view. They are detectable
mostly in red/infra-red wavelengths and only for about 1 or 2 weeks in an instrument such as DECam. With only one kilonova associated to a GW event so far, light-curve models
\citep{2012ApJ...746...48M,barnes2013,2013ApJ...775..113T,grossman2014,2017LRR....20....3M,2017CQGra..34j4001R,2018ApJ...869..130R,2018ApJ...860...62G,2018ApJ...869L...3F,2019arXiv190604205B} are still very uncertain. They predict, for example, a wide range of peak magnitudes: e.g.~$r$-band (650nm)
magnitude 20--22 at a distance of 200 Mpc.

\subsection{Precision Cosmology}

$H_0$ is a powerful tool for precision cosmology which aims at characterizing the Universe as a whole: physics at the early times, history
of expansion, and behavior of perturbations. With a distance-based 
measurement, we probe the history of expansion and are sensitive to dark
energy models (often described in terms of the equation of state parameter 
$w=p/\rho$). 
Discrepancies between $H_0$ values determined from the Cosmic Microwave Background and from distance
probes can be reconciled by allowing some of these model parameters to vary.

Uncertainties in $H_0$ measurements from standard sirens will depend, in part, on
uncertainties achieved by the LVC network for the distance to each merger. 
According to a study~\citep{2013arXiv1307.2638N} of simulated binary neutron star mergers, with a 
distance distribution and detector performance model that are representative of the third and fourth observing runs, 
an uncertainty of 3--5\%\ on $H_0$ is expected for sample sizes of about 15--20 events. 
Similarly, for {\it dark sirens}, or GW events without
an identified electromagnetic counterpart where the redshift must be obtained statistically~\citep{2008PhRvD..77d3512M},
simulations following the methodology of \cite{Soares_Santos_2019} estimate that 5\%\ precision is achievable 
with a sample of 100 events (see also \citealt{2018Natur.562..545C} and \citealt{2018PhRvD..98b3502N} for other estimates of the achievable precision on $H_0$ via standard sirens).
After approximately 100 dark siren events and 20 bright siren events, we expect that our program can contribute to
the $H_0$ discrepancy issue 
with a completely independent measurement at the $2\sigma$ level, a remarkable 
result for a brand new cosmological tool.\footnote{\citet{2020arXiv200504325P} propose to use gravitational wave compact binary mergers and large galaxy surveys to also set constraints on gravity and the growth of structure by measuring their peculiar velocity power spectrum, overdensity, and cross--correlation power spectra.}

\section{The DECam System}\label{sec:decam}

We use the Dark Energy Camera~\citep{flaugher2015}. 
DECam is a 570 Mpixel, 3-square-degree field-of-view camera with an active-pixel fill-factor of 0.8. It consists of  
62 red-sensitive CCDs and
a wide field of view corrector, all installed on the 4-meter Victor M. Blanco telescope at the Cerro Tololo 
Inter-American Observatory (CTIO).  
This sensor system is connected to a 
data reduction pipeline with 
computation hardware at the National Center
for Supercomputing Applications (NCSA).

\subsection{The DES, DECam, and the Community}

The Dark Energy Survey Collaboration (DES) designed and constructed DECam with the primary goal of studying the nature of dark 
energy using four complementary probes:  galaxy clusters, weak lensing, 
Type Ia supernovae, and baryon acoustic oscillations. 
As we have seen, precision measurements of $H_0$ are intricately linked with dark energy itself, thus 
our GW program complements the main DES probes.
The camera was installed in 2012, significantly upgrading the Blanco
telescope, completed in 1976.
The high level requirements on the DECam design were driven by 
the need to visit a 5000 sq. deg. wide-survey area (the DES footprint) and a 30 sq. deg. supernova survey area over five years, 
with excellent image quality, high sensitivity 
in the near infrared, and low readout noise.

\subsection{Template Coverage}

\begin{figure}[htp]
\centering
\includegraphics[width=0.95\linewidth]{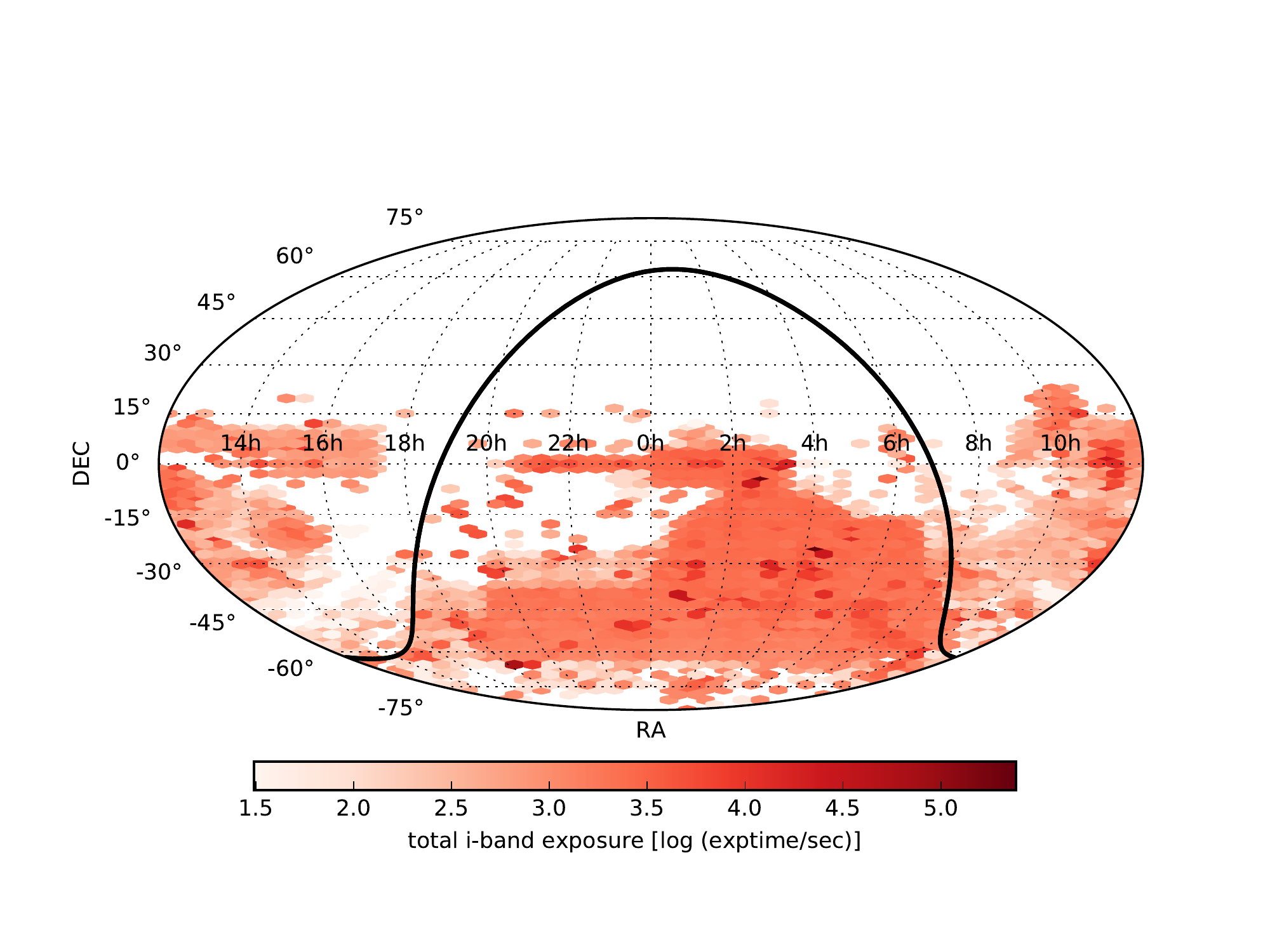}
\caption{Existing $i$-band coverage of the sky from DECam imaging as of March 2019.
\label{fig:imap} }
\end{figure}

Our transient discovery pipeline uses the \ldquo image differencing\rdquo  technique. This technique works by comparing a search image to one or 
more images previously taken of the same area of sky (these images are called ``templates")
and subtracting the template(s) from the search image. The result, known as the ``difference image", should consist
only of objects not present in the template images, which are thus potential candidates to be an optical counterpart of the GW trigger. Inside the DES footprint, template images
are always available, either from the DES wide-survey program or from other programs with publicly accessible data. Outside the DES footprint we rely on public images taken by
other programs using DECam to provide templates. This method results in 
incomplete sky coverage (e.g. Fig.~\ref{fig:imap}), illustrating
the need for programs to complete the southern sky
coverage, such as the Blanco Imaging of the Southern Sky (BLISS) survey. 
We require a minimum exposure time of 30 seconds in our templates, and our standard 
observing conditions requirements translate to effective magnitudes of 21.23 (20.57) for 30-second $i$ ($z$)-band images. Currently DECam public image coverage to this depth exists over approximately
59\% (65\%) of the accessible sky in the $i$ ($z$) band. If no template images pass the standard observing criteria we can loosen them as needed (though we retain the 30-second minimum) at the expense of
search sensitivity (we are limited by the combined template depth), resulting in increased coverage.
If no DECam template image was available at a given location
we relied on so-called ``late-time" images taken well after the event
as templates. Late-time templates are non-optimal for two reasons: first, any residual light
from the event in the late-time image systematically affects measured magnitudes of the event
at earlier times. Second, the use of late-time template delays complete spectroscopic response
until after the template is taken. 

\subsection{How the Interrupt Process Worked During DES Time}
The first two LIGO observing runs coincided almost exactly with DES observing seasons.
The DESGW group, consisting of members of
the DES, LVC, and other members of the astronomical community, 
was awarded $\sim 4$ nights of telescope time per year via
the open proposal process. DES and the DESGW entered into an agreement
where the DESGW nights were added to the DES observing time allocation. 
If a trigger occurred, the DESGW and DES management teams consulted and if the event
was judged interesting the DESGW interrupted the DES and made observations of the LIGO spatial localization. If the time
was not used, the nights counted against the total night allocation of the
DES.

\section{The Experimental Setup}\label{sec:exsetup}

\begin{figure*}[!htb]
\includegraphics[trim={0 0 0 0cm},clip,width=\linewidth]{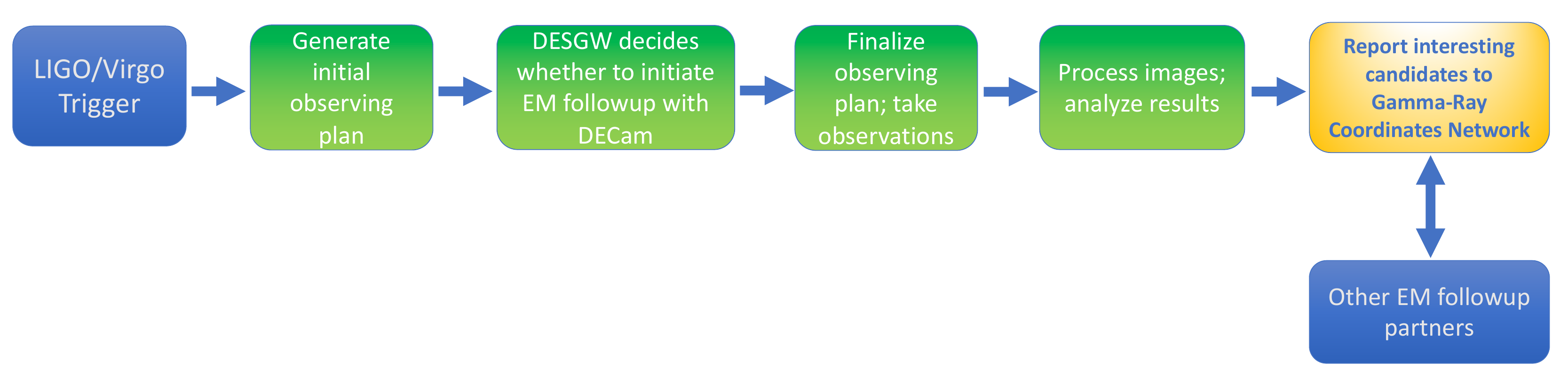}
\caption{Overview of search program (green boxes are specific to this work).
 Our automated system continuously listens to the LIGO/Virgo
 data stream via GCN protocol. For a given trigger,
 an automated strategy code formulates the observing
 plan which we send to astronomers at the DES/DECam
 site to execute. We monitor the DECam data stream
 for new incoming images and automatically process them through the
 image differencing} pipeline.
\label{searchprogram}
\end{figure*}

Figure~\ref{searchprogram} provides an overview of the experimental setup.
The reception of a GCN trigger from LIGO triggered a three-part DESGW process.
First, we calculated the possible coverage of the LIGO event with the DECam system
for the next night and discussed whether to proceed with observations as previously described.
If the decision was to interrupt and observe, we moved to the observation and data preparation phase.
The observing team was briefed and provided the observing plan, and the template determination and preparation began.
We make an initial determination of possible template overlaps based on our observing plan, discussed in detail in Sec.~\ref{sec:mapmaking}.

The third phase is data reduction. 
During O1 and O2, the DESGW program aimed to process and analyze
a given night's observations within 24 hours so that we can report electromagnetic 
counterpart candidates to other telescopes for followup. 
We rapidly provision the required computing resources using
a mixture of computing resources at Fermilab and a variety of other campus and laboratory sites via the Open Science Grid (OSG; ~\citealt{Pordes_2007}), relying on the high throughput of the OSG.
As new search images arrive at Fermilab from CTIO via NCSA,
the corresponding image processing jobs are sent to all available resources.
The results are copied to local disk at Fermilab and summary data recorded in a database for postprocessing
and candidate selection.

\section{DESGW Observing: Map Making and Observing}\label{sec:mapmaking}

Our system listens to the LIGO/Virgo alert stream. The MainInjector code
initiates action on triggers by alerting DESGW team members and preparing
initial observation maps. The observing scripts are generated by calculating
event counterpart visibility probability maps summed inside an all-sky DECam
hex layout and choosing the highest-probability hexes for highest priority observations.

We wish to calculate maps of
\begin{equation}
p(\alpha,\delta) = 
\int p(x|\alpha, \delta)
p(y|\alpha,\delta)  
p(d|m,t,\alpha,\delta)
dm \,.
\end{equation}
where 
\begin{itemize}
\item $p(x|\alpha,\delta)$ is the LIGO spatial localization map probability per pixel ($x$),
\item $p(y|\alpha,\delta)$ is our ability to recognize the detection given source crowding and is
related to a false positive rate, 
\item $p(d|m,t,\alpha,\delta)$ is the DECam detection probability of a source of magnitude $m$
at time $t$ per pixel, and
\item the time dependence of $m$ enters via the source model.
\end{itemize}
We will discuss each of these items in turn.

\begin{figure*}[!htp]
\centering
\subfigure[][DES source probability time slot 1]{\includegraphics[width=0.32\textwidth]{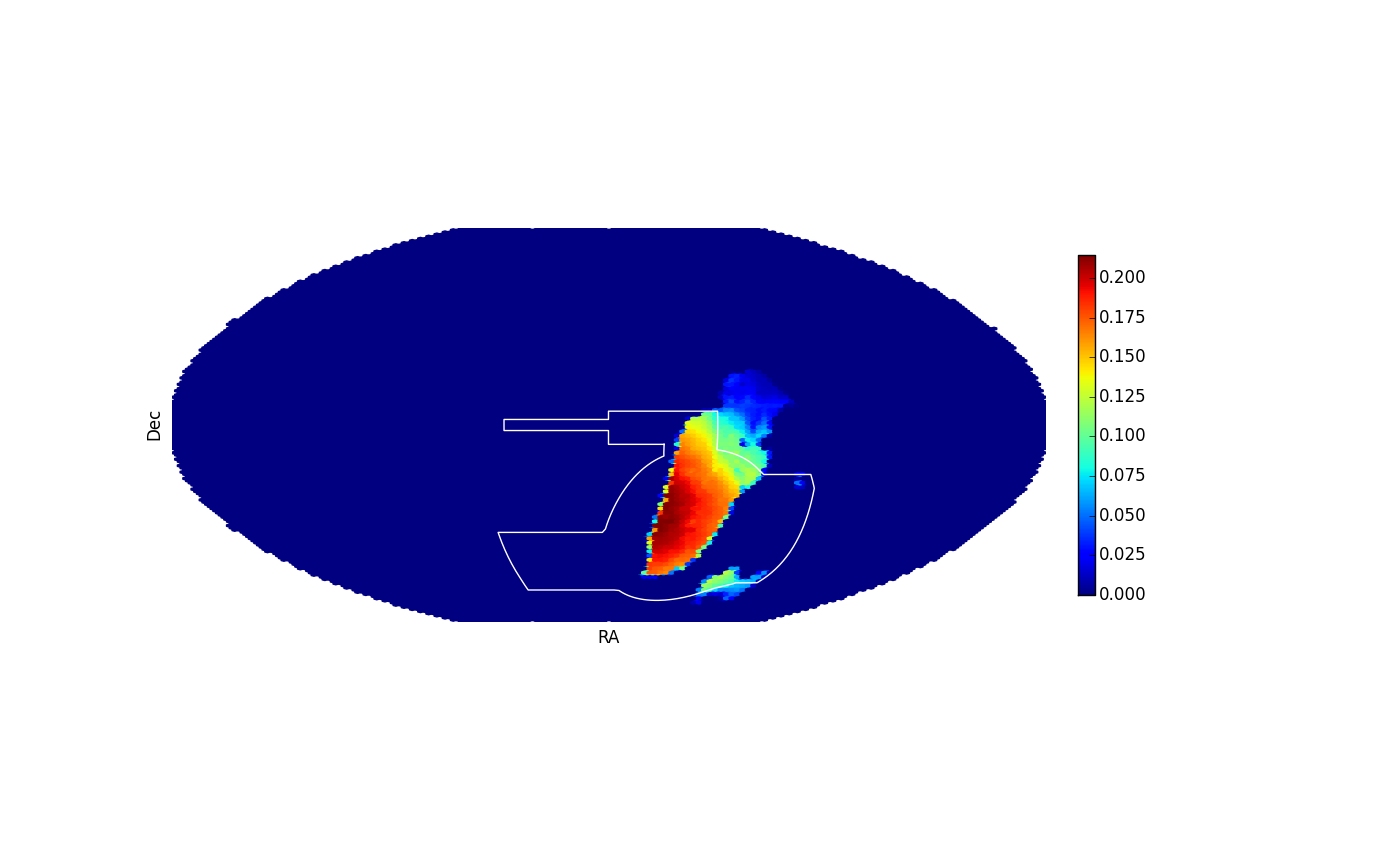}\label{desprob1}}
\subfigure[][Limiting magnitude map time slot 1]{\includegraphics[width=0.32\textwidth]{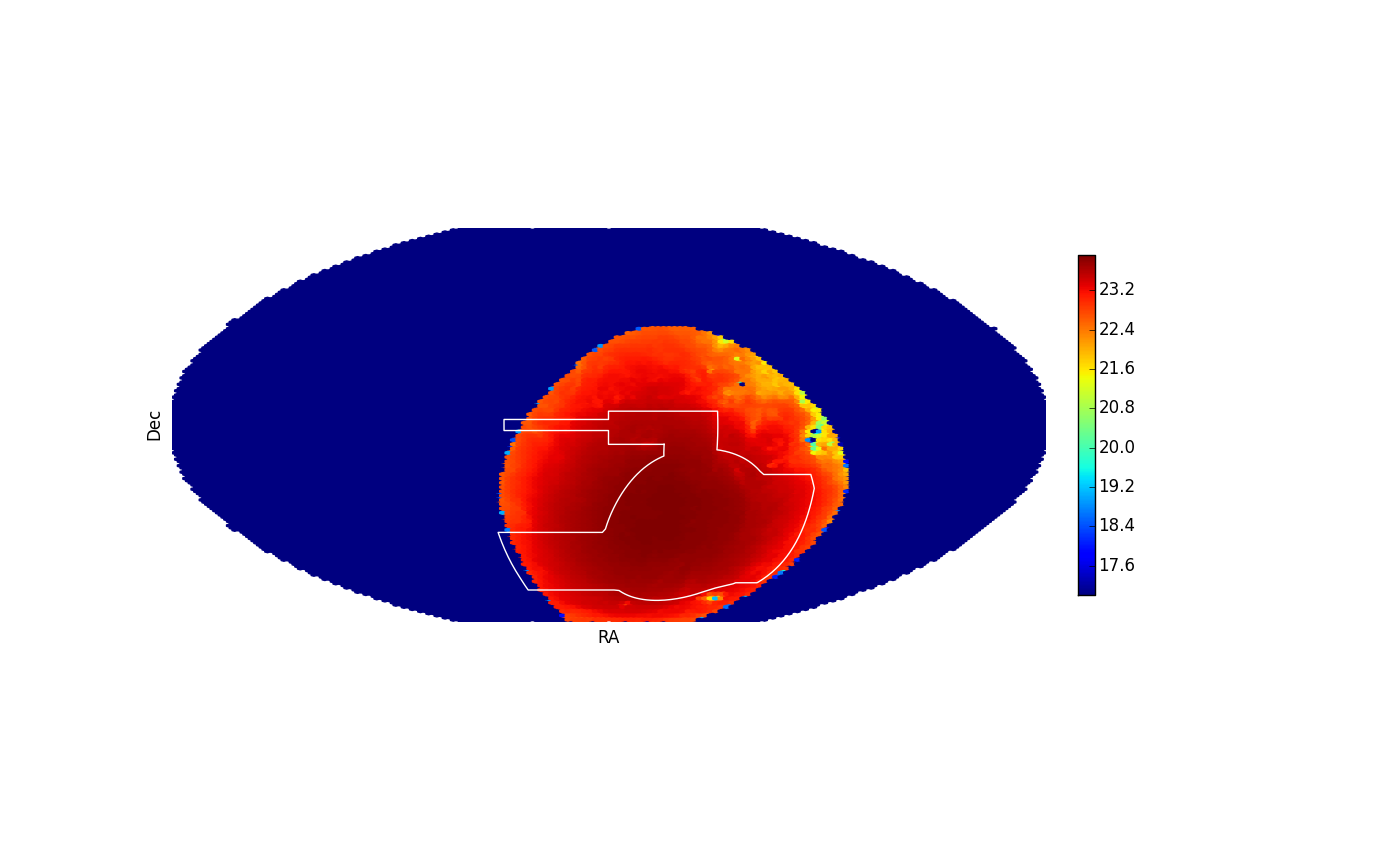}\label{maglim1}}
\subfigure[][DES $\times$ LIGO probability time slot 1]{\includegraphics[width=0.32\textwidth]{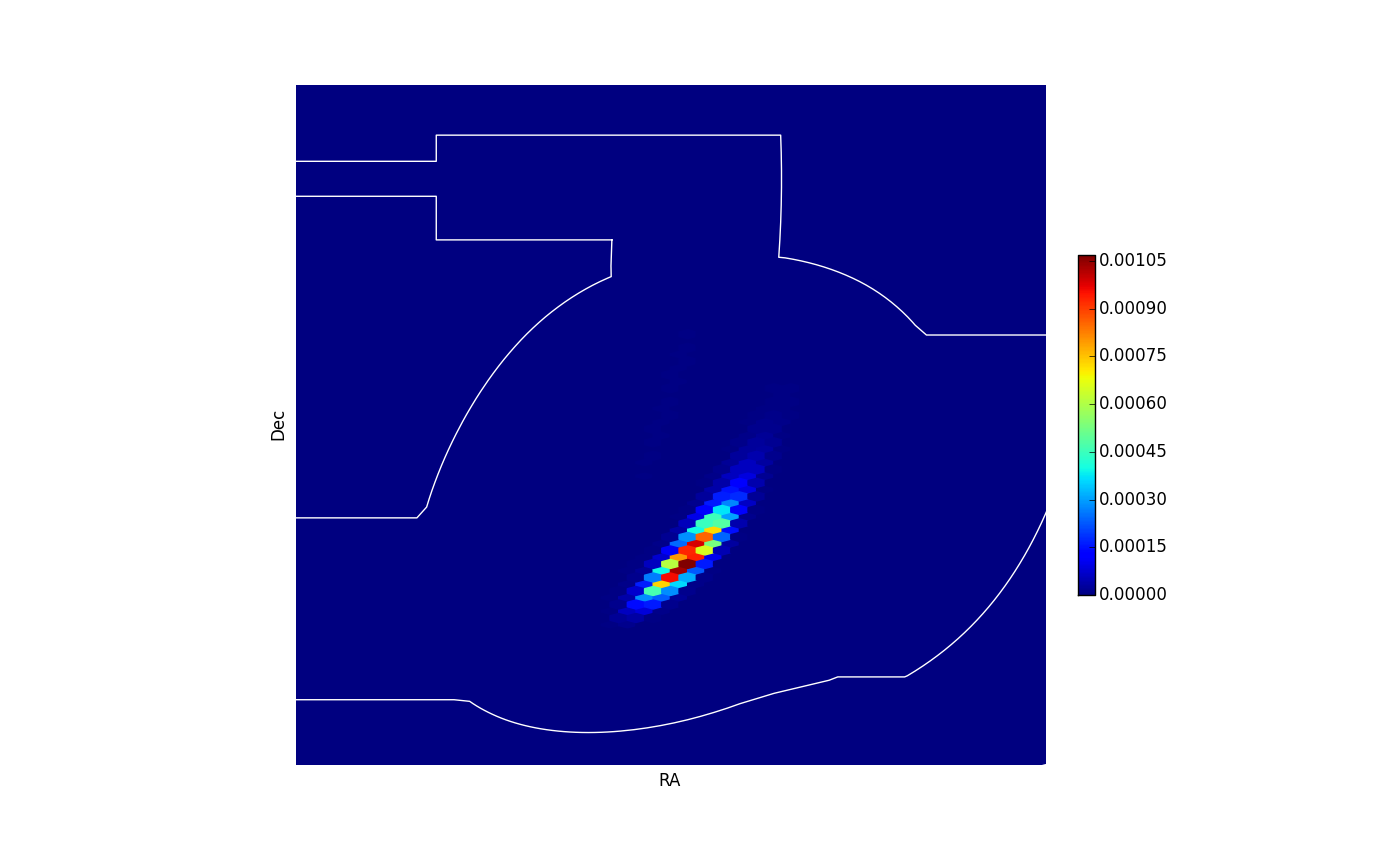}\label{ligoxdes1}}\\
\subfigure[][DES source probability time slot 2]{\includegraphics[width=0.32\textwidth]{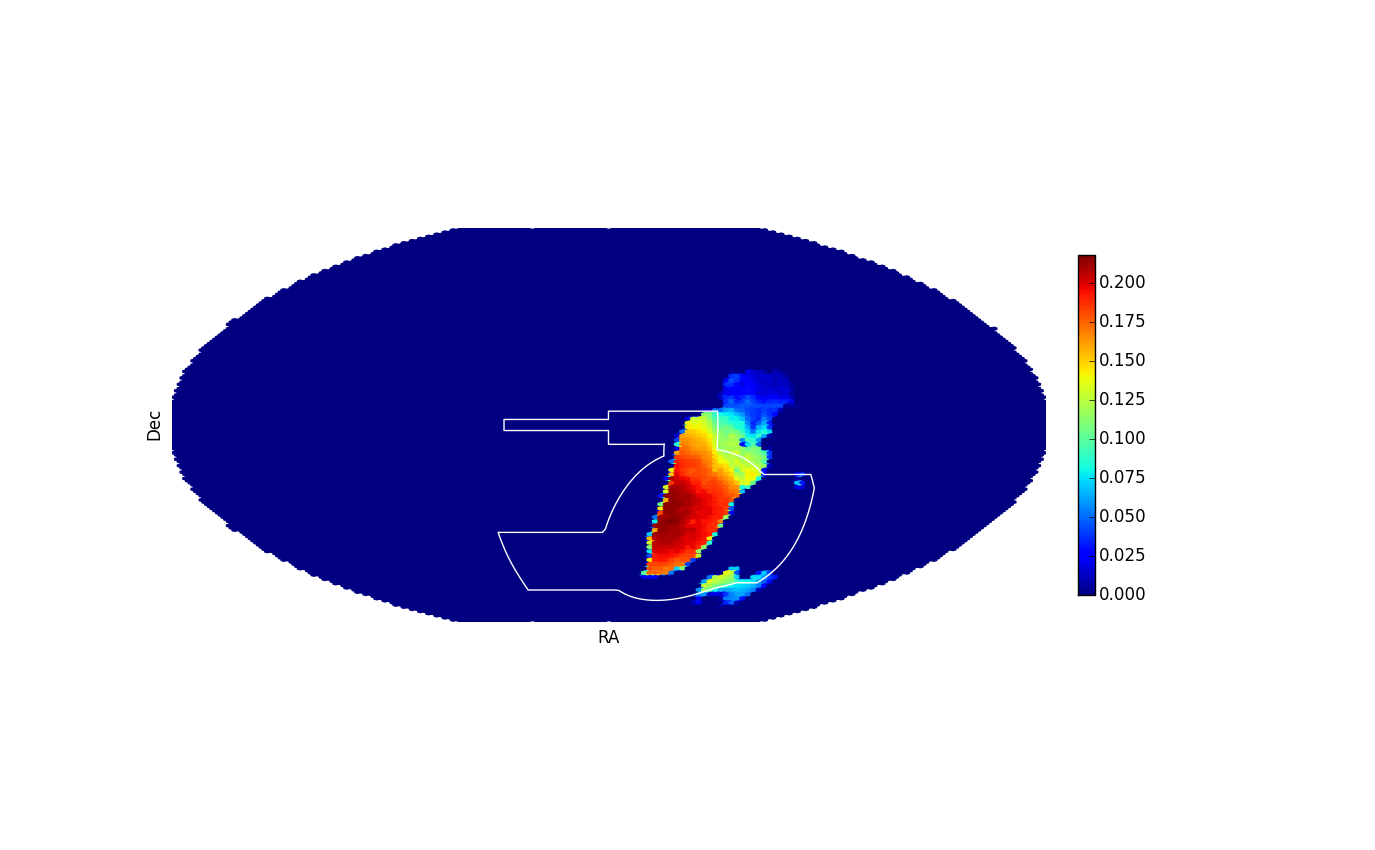}\label{desprob2}}
\subfigure[][Limiting magnitude map time slot 2]{\includegraphics[width=0.32\textwidth]{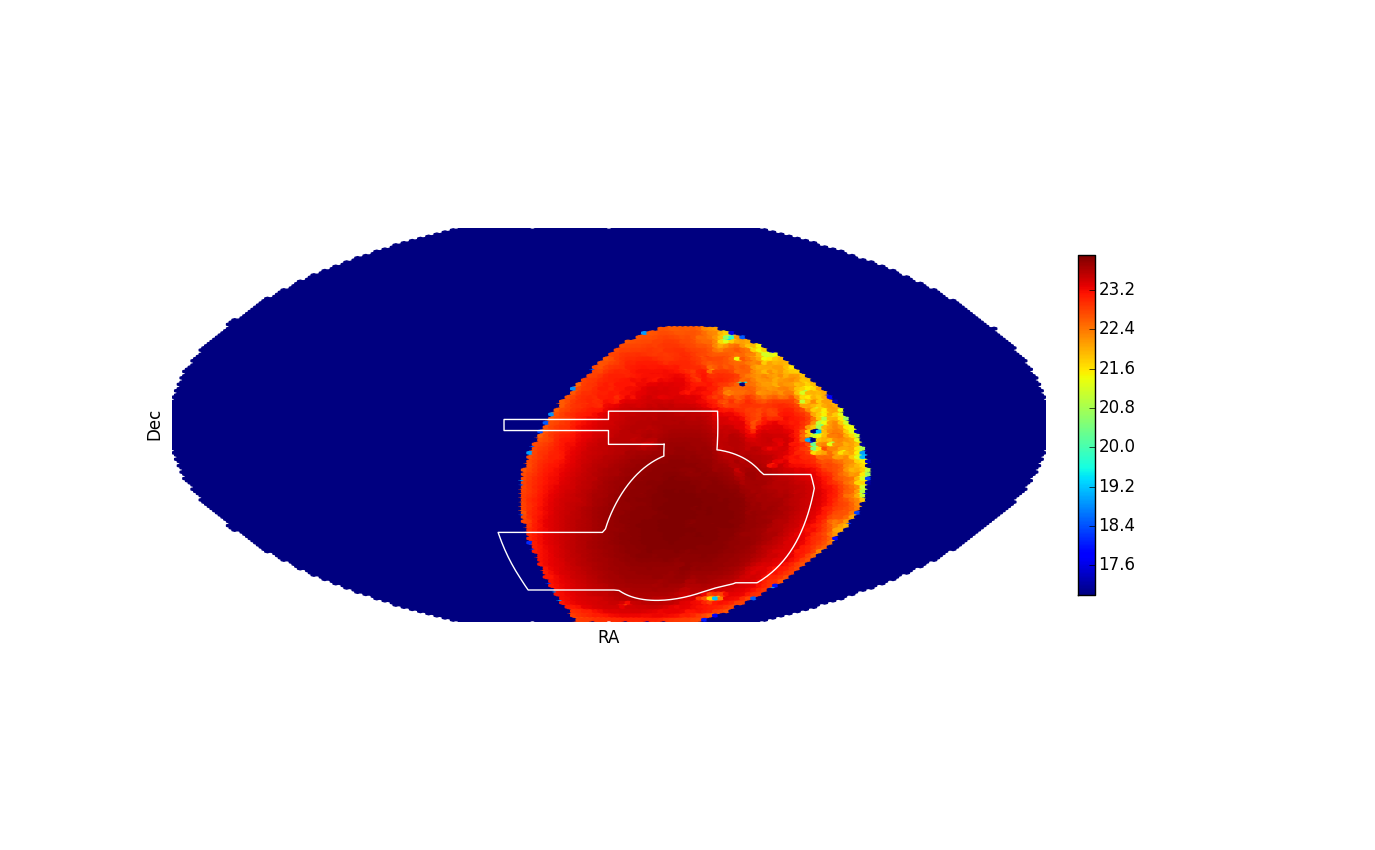}\label{maglim2}}
\subfigure[][DES $\times$ LIGO probability time slot 2]{\includegraphics[width=0.32\textwidth]{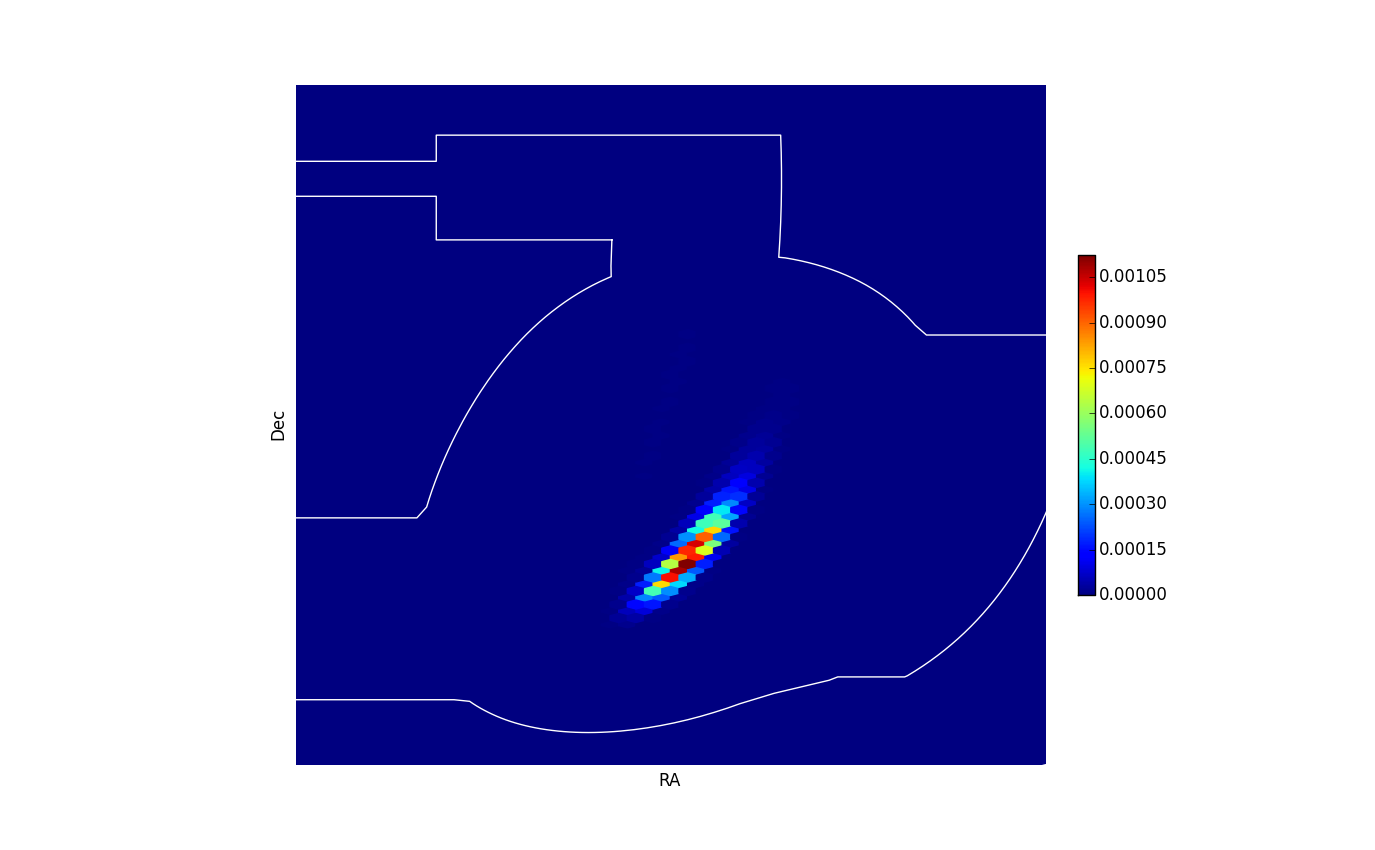}\label{ligoxdes2}}\\
\subfigure[][DES source probability time slot 3]{\includegraphics[width=0.32\textwidth]{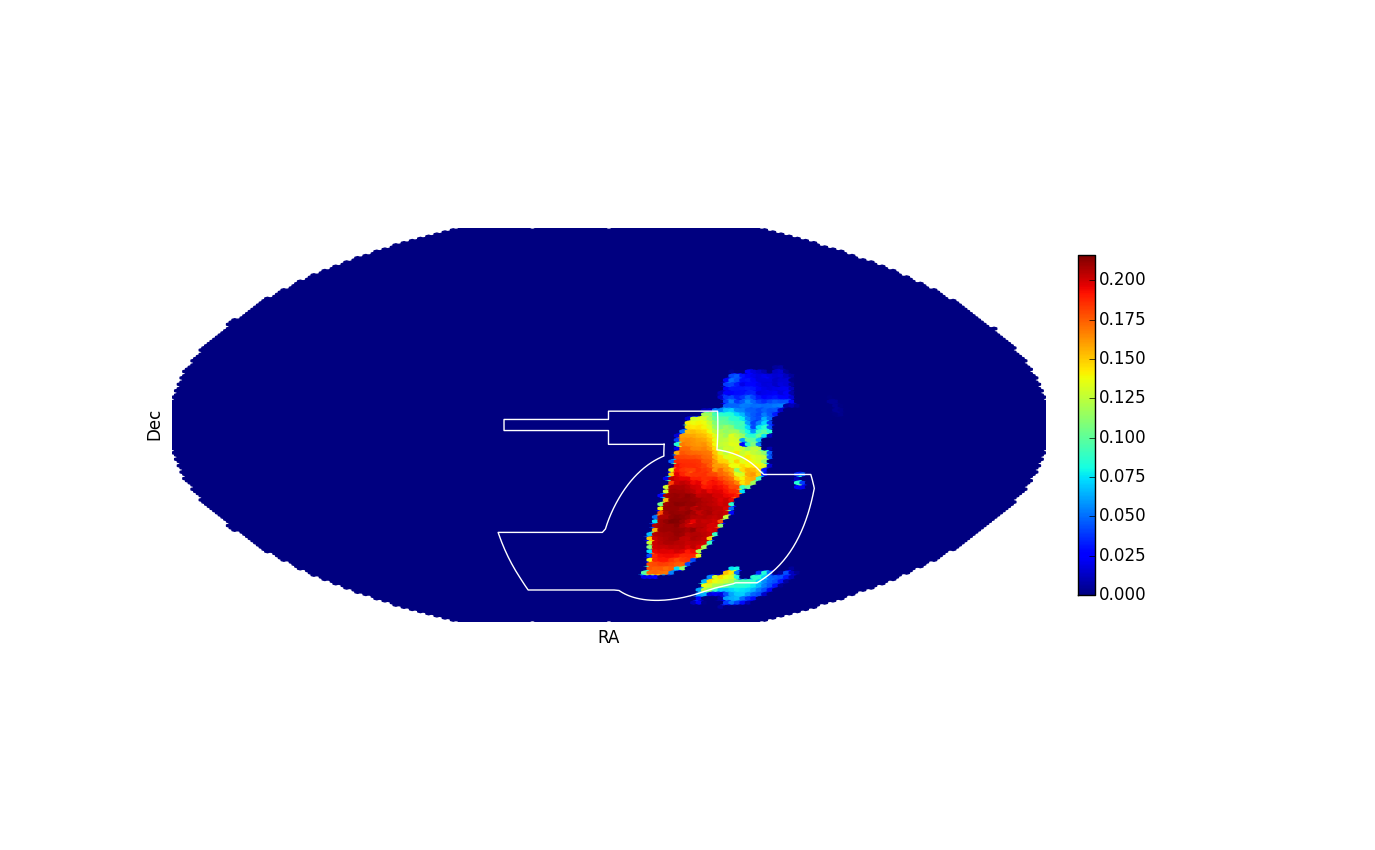}\label{desprob3}}
\subfigure[][Limiting magnitude map time slot 3]{\includegraphics[width=0.32\textwidth]{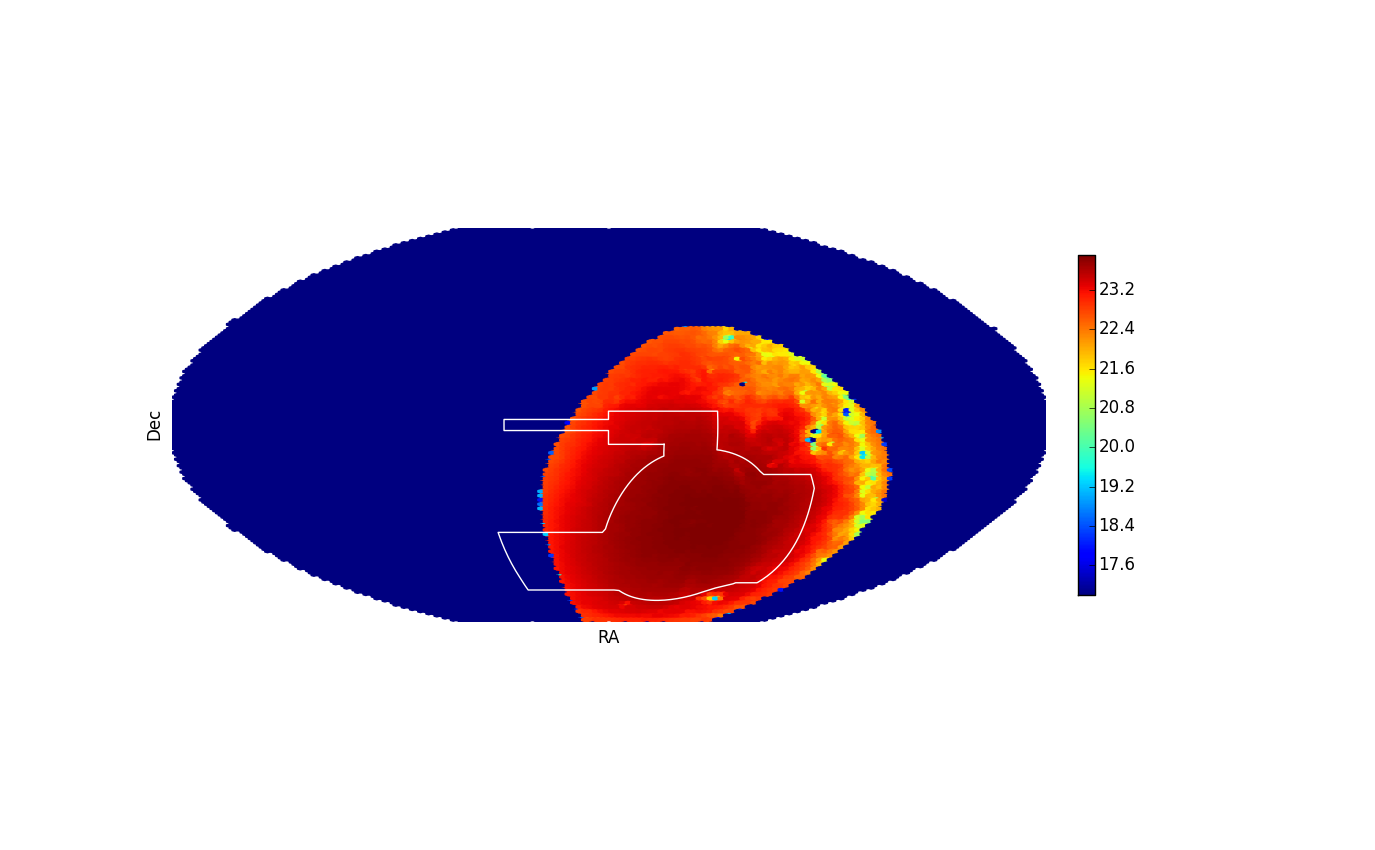}\label{maglim3}}
\subfigure[][DES $\times$ LIGO probability time slot 3]{\includegraphics[width=0.32\textwidth]{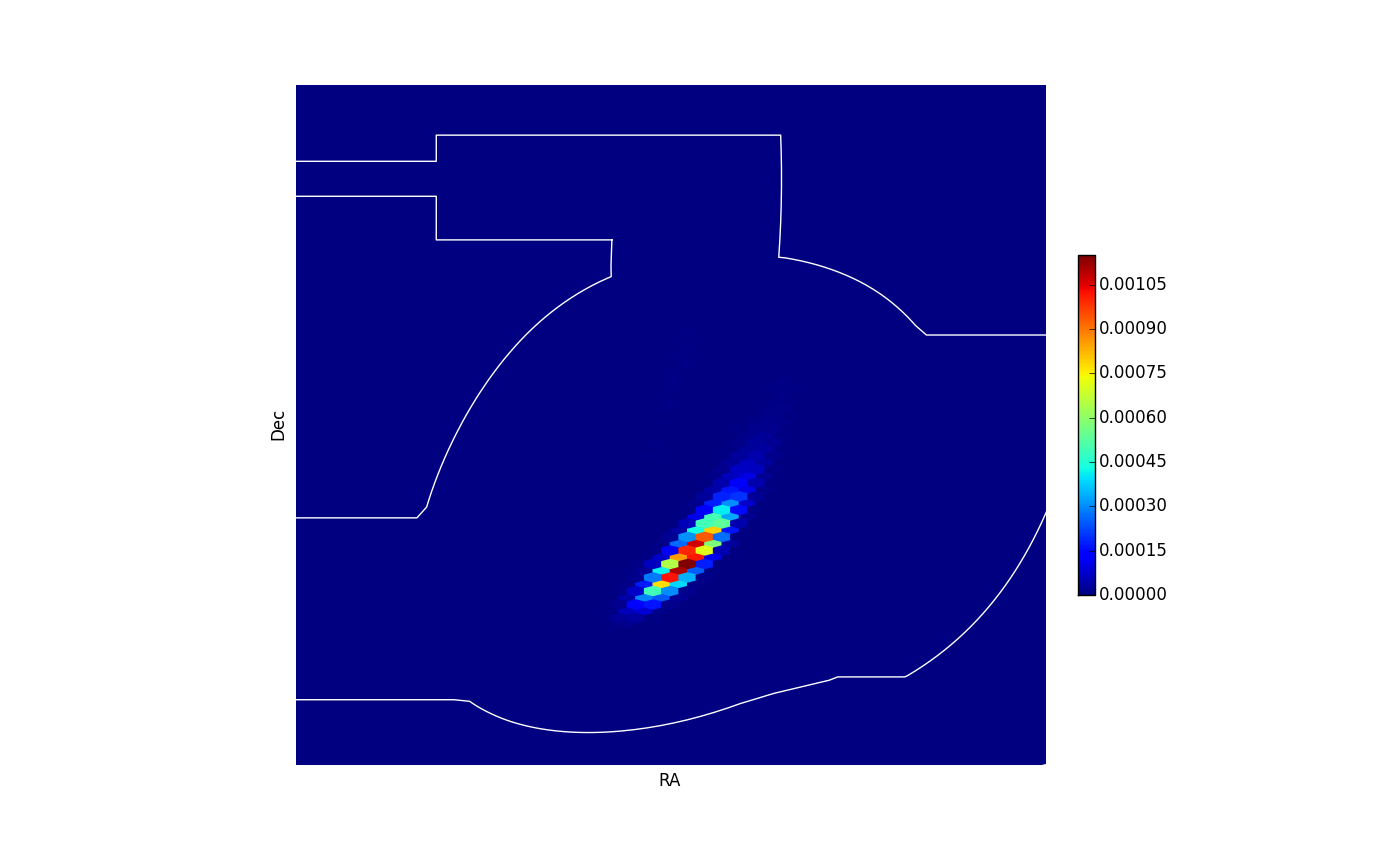}\label{ligoxdes3}}
\caption{Source detection probability map construction.\label{fig:mapmatrix}. We show the relevant probabilities and maps for three different time periods on a given night. While the source probability map is fixed (\subref{desprob1},\subref{desprob2},\subref{desprob3}), the limiting magnitude map (\subref{maglim1},\subref{maglim2},\subref{maglim3}), and thus the overall source detection probability map (\subref{ligoxdes1},\subref{ligoxdes2},\subref{ligoxdes3}), changes with time due to observing conditions and the Earth's rotation. The example shown here is for GW170814, a binary black hole merger. The white contour represents the DES footprint.}
\end{figure*}

\subsection{Spatial Localization: $p(x|\alpha,\delta)$}

LVC triggers include localization maps in HEALPix format (LIGO maps: 
\citealt{Singer2016}, HEALPix: \citealt{2005ApJ...622..759G}). 
The first of these maps provides the spatial localization probability per pixel, the second provides distance,
the third provides a gaussian variance estimate on the distance, and the fourth contains a normalization plane.
We take the first map to be 
$p(x|\alpha,\delta)$. Maps derived from this one have the same resolution as we choose to read this map.

\subsection{Detection recognition probability map: $p(y|\alpha,\delta)$ }

We model $p(y|\alpha,\delta)$, the probability of recognizing a detection $y$ given source crowding, as a false positive rate problem related to the per-pixel stellar density and independent of magnitude.
Essentially, if the targeted region is too full of stars then it is likely there will be too many false positives masking possible candidates, resulting in 
missed detections of real astrophysical transients.
Our model is that the recognition probability is taken to be $p(y|\alpha,\delta)$ = 0.0 at or above 610 stars/deg$^2$ (roughly that of the Galactic anticenter) and $p(y|\alpha,\delta)$ = 1.0 at or below 10 stars/deg$^2$ (roughly that of the south Galactic pole), linear in surface density between.  
We use the 2MASS J-band star counts in units of stars/deg$^2$, cut at $J < 16.0$, from 
the NOMAD catalog~\citep{2004AAS...205.4815Z}.
Figures~\ref{desprob1}, \ref{desprob2}, and \ref{desprob3} show
example detection recognition probability maps for a given event.

\subsection{Limiting magnitude maps}\label{subsec:maglim}

The calculation of $p(d|m,t,\alpha,\delta)$ proceeds through limiting magnitude maps,
some examples of which are shown in Figs.~\ref{maglim1}, \ref{maglim2}, and \ref{maglim3}.
A point source with intrinsic flux $f$ and full width half max $w$ that is faint
compared to a sky background flux $s$ has a signal to noise S/N given by
\begin{equation}\label{signal-to-noise}
S/N \propto  \frac{T_d T_a f t}{w \sqrt{s t}} = \frac{T_d T_a f \sqrt{t}} {w\sqrt{s}}
\end{equation}
where $T_d$ is the transmission of light through a layer of Galactic dust,
and $T_a$ is the transmission of light through the Earth's atmosphere.
We demand S/N=10 for a detection as $10\%$ flux uncertainties permit reasonable colors to be formed.

We define a fiducial set of conditions and a 90 sec DECam exposure
to calculate the $10\sigma$ $i$-band magnitude, $m_i$, 
and use eq~\ref{signal-to-noise} to compute
the limiting magnitude $m_l$ using
$\Delta m$ to account for variations from the fiducial values:
\begin{align}
\begin{split}
 m_l &= m_i + \Delta m\\
\Delta m &= 
         2.5\log\left[ \left(\frac{T_d}{T_{ncp}}\right) \left(\frac{T_a}{T_{1.3}}\right) 
         \left(\frac{0.9\arcsec}{{\mathrm FWHM}}\right)\right] \\
         &+ \frac{1}{2}(m_{sky} -22.0)
\end{split}
\end{align}
where $T_{ncp}$ is the transmission of dust at the North Celestial Pole,
and $T_{1.3}$ is the transmission of the atmosphere at an airmass of 1.3, our fiducial value.
This quantity is entirely a function of time of the observation: 
$T_a$ depends on zenith distance; the full width of half maximum of the PSF, $\mathrm{FWHM}$,  
depends on zenith distance;
$m_{sky}$ depends on zenith distance,  the 
lunar phase and position, and the solar position.

{\sc Transmission due to Galactic dust:}
The Planck dust map comes as a HEALPix map, with the values of the pixels being
the optical depth $\tau$ at 350 $\mu$m multiplied by a normalization to convert to $E(B-V)$, 
the reddening.
This is best understood as $A_B-A_V$, where the $A$ is the extinction
in a given filter. The extinction in the $i$-band, $A_i$, is therefore $R_i \times E(B-V)$,
where $R_i$ is the reddening coefficient for the $i$-band.
The transmission due to dust, $T_d$, is $\log_{10}(T_d) = 0.4 \cdot R_i \cdot E(B-V)$
so the transmission relative to the fiducial is thus
\begin{align}
\begin{split}
\log_{10}(T_d) &= 0.4 \cdot R_i \cdot E(B-V) \\
\log_{10}(T_{ncp}) &= 0.4 \cdot R_i \cdot E(B-V)_{ncp} \\
\log_{10}\left(\frac{T_d}{T_{ncp}}\right) &= 0.4  R_i [E(B-V) - E(B-V)_{ncp}]
\end{split}
\end{align}

We take $E(B-V)_{ncp}$ to be 0.0041.

{\sc Atmospheric transmission:}
In a very simple model the transmission due to the atmosphere can be described as
\begin{equation}
\frac{T_a}{T_{a,f}} = \frac{10^{-0.4 k_i X }}{10^{-0.4\cdot 1.3 k_i}} 
= 10^{-0.4 k_i \left(X-1.3\right)}
\end{equation}
where $k_i$ is the (first order) atmospheric extinction in the $i$-band,
and $X$ is the airmass.
Since zenith distance could be close to $90^\circ$ in our case, depending on
the target of interest, we do not 
use the $X=\sec(zd)$ approximation to airmass, but the
approximation due to~\cite{1994ApOpt..33.1108Y}, which has a maximum error of 0.004
at $zd=90^\circ$.
\begin{align}
    X = \frac{a\cos^2(zd) + b\cos(zd) + c}
        { \cos^3(zd) + d\cos^2(zd) + e\cos(zd) + f}
\end{align}
where $a=1.002432$, $b=0.148386$, $c=0.0096467$, $d=0.149864$, $e=0.0102963$,
and $f=0.000303978$.

{\sc The full width half max of the psf:}
$\mathrm{FWHM}$ is, statistically,  a filter-dependent  power law in zenith distance.
\begin{equation}
\frac{\mathrm{FWHM}}{\mathrm{FWHM}_f} = \left(\frac{\lambda}{\lambda_f}\right)^{-0.2} \left(\frac{X}{1.3}\right)^{3/5}
\end{equation}

{\sc The sky surface brightness:}
The sky surface brightness model is from \cite{krisciunas1991},
which predicts the sky brightness as a function of
the moon's phase and zenith distance, the zenith distance of the sky position, 
the angular separation of the moon and sky position, the local extinction coefficient,
and the airmass of the sky position.
Telescopes have software and hardware limits on their ability to point, which we model
as a top hat on sky brightness: outside the pointing range the
sky brightness is set arbitrarily high (the brightness of the full moon).
We create this brightness map by converting the CTIO-provided Blanco HA,$\delta$ telescope limits into a HA,$\delta$  map. 

\subsection{Source detection probability: $p(d|m,t,\alpha,\delta)$}

The calculation of the
map $p(d|m,t,\alpha,\delta)$ uses the limiting magnitude map,
the source model, and the LIGO event distance map.
Discussion of the source model is in the next subsection and here we assume that model gives us an absolute magnitude
$M$ and a model uncertainty dispersion of $\sigma_M$.
We pursue the following calculation.

We convert the distance estimate from the LIGO trigger into a 
distance modulus 
$\mu = 5 \log r + 25$, 
where the distance $r$ is in Mpc,
and the distance uncertainty into an uncertainty on $\mu$:
\begin{equation}
\sigma^2 = \sigma_r^2 \left(\frac{5}{r\ln 10}\right)^2
\end{equation}
Then the apparent magnitude of the source, 
$m_s$ is
\begin{equation}
m_s = M + \mu
\end{equation}
and, as gaussians can be added to form gaussians,
\begin{equation}
\sigma_m^2 = \sigma_M^2 + \left(\frac{5}{r \ln 10}\right)^2 \sigma_r^2
\end{equation}
The resulting gaussian is normalized:
$\int_0^\infty G(m_s, \sigma_m^2) dm = 1$.
The probability that the object could be seen in this
pixel given a limiting magnitude of $m_l$ is then
\begin{equation}
p(d|m_l, m_s,\sigma_m) = \int_0^{m_l} G(m_s,\sigma_m^2) dm
\end{equation}
If $m_s > m_l$, then $p(d) = {\mathrm{efc}}(m_s-m_l)/2$.
If $m_s < m_l$, then $p(d) = (1 +{\mathrm{efn}}(m_s-m_l))/2$.
The probability is weighted as uniform in volume: $\int p(d) dr = \int G(m_s,\sigma_m^2)r^2 dr$.
We will need to do the same weighting in apparent magnitude space,
by transforming $r^2$ into distance modulus
\begin{equation}
r = 10^{0.2(m-m_s) - 5} = 10^{0.2(m-m_s) +C}
\end{equation}
\begin{equation}
r^2 dr = C\, 10^{0.6 (m-m_s)} dm 
\end{equation}
where $C$ is a constant.
Then

\begin{equation}
\int_0^r  G r^2 dr = C \int_0^{m_l} \exp\left(\frac{(m-m_s)^2}{\sigma_m^2}\right) 
10^{0.6 (m-m_s)} dm
\end{equation}
where $C$ is a normalization constant.
This last equation is the probability that we will see a source given
the model uncertainty and distance uncertainty for a known limiting magnitude 
and thus is
$p(d|m,t,\alpha,\delta)$: recall $d$ here is a detection,
$m_l$ is a function of spatial pixel ($\alpha,\delta$) and time ($t$) as described in Sec.~\ref{subsec:maglim},
and $m_s$ depends on $M$, its uncertainty, and time as described in the next subsection.

\subsection{The source model}

The source model governs the source magnitude and uncertainty as a function of time.
We operate with two theoretical source models, one for binary black hole mergers and
one for mergers including a neutron star.

Our source model for BBH mergers is relatively arbitrary, as
there is no well-motivated theoretical model. We assume simply that the counterpart has
a falling $t^{-1}$ lightcurve: the source has an initial $i$-band magnitude of approximately 20 for a day or two, then
fades below our typical detection thresholds for 90-sec exposures.
Our source model for a merger involving a neutron star is more complicated.
In O1 and O2, before detection of the BNS event GW170817, we were following the kilonova model of \cite{barnes2013} and \cite{2016ApJ...829..110B}. 

Briefly, as neutron stars merge,
tidal tails grow into the equipotential surface and then mostly flow
back into the main remnant. Some few percent of matter in the tails are dynamically ejected
at speeds $\approx$  0.1c. 
The ejected material undergoes r-process nucleosyntheis 
as the few nuclei in the sea of neutrons grow by neutron absorption. 
During O1 and O2 the details of our model followed the simulation-based
analysis of \cite{grossman2014}. 
Observationally, we modeled the merger event luminosity, $L$,
and temperature, $T$, and assumed the event is an optically thick
blackbody of which we observe a photosphere. All the thermal energy
available for comes from the
the r-process beta-decay episode and 
we define an energy deposition rate per unit mass, $\epsilon$,
so that $L \approx m\epsilon$.
At early times ($t < t_{peak}$) $L$ scales as 
$L \propto t^2$ and at late times as ($t > t_{peak}$) $L \propto t^{-1/3}$.
The time to peak brightness $t_p$ is $\sim m^{-1/2}$.
We assume the photosphere is a blackbody
$T = \left(  \frac{L}{\sigma \left(2\sqrt{\pi} v t \right)^{2}} \right)^{0.25} $,
where $vt$ came from the radius of the photosphere.
The flux through any filter is then:
$  f= 4.4 \times 10^{22}\,\, \int B_\lambda {\rm d}\lambda\,\, 
        L_{40} T_{1000}^{-4} d_{100}^{-2}$ 
where  $L_{40}$ is the luminosity in units of $10^{40}$ ergs/sec,
$T_{1000}$ is the temperature in units of $1000$ K, and $d_{100}$ is the
distance in units of 100 Mpc.

We can break up the model into sub-components, each of which has a different
opacity. In our model we use a Barnes and Kasen iron model $(\kappa \approx\ 0.1)$
and an Barnes and Kasen lanthanide model $(\kappa \approx\ 10)$.
We use the same energetics as described
above; one swaps out the $B_\lambda$ for a computed SED as one computes
the flux through a filter. 

There was a range of uncertainties in the 
peak absolute magnitudes.
For a given time we calculate the model absolute magnitude,
$M$ which has a model uncertainty dispersion of $\sigma_M$.

During and after the intensive observations of the counterpart to GW170817, better models have been developed, for example the three-component model of \citet{2017ApJ...851L..21V}. It was unclear whether there would be a blue component before the observations of  GW170817; in the work described here we assumed only a red component and a generally fainter absolute magnitude than the counterpart of GW1701817 exhibited. We have updated the kilonova model implementation in our O3 pipeline.

\subsection{Observing plan construction}

Once we have calculated maps of $ p(\alpha,\delta)$
we can construct the observing plan. Each observation is of a hex,
so named because the DECam focal plane is roughly hexagonal.
An observation may be a single 90s $i$-band exposure, as in BBH sources,
or a triplet of 90s each in an $i,z,z$ band sequence, as in events containing NS.
The aim is different in each
case. For NS we are looking for the week-timescale very red transient that
is the signature of kilonova. For BH we are assuming a source similar to
a $t^{-1}$ power law and attempting to build light curves that allow
us to reject supernovae.

\begin{figure*}[htb!]
\centering
\includegraphics[width=0.33\textwidth]{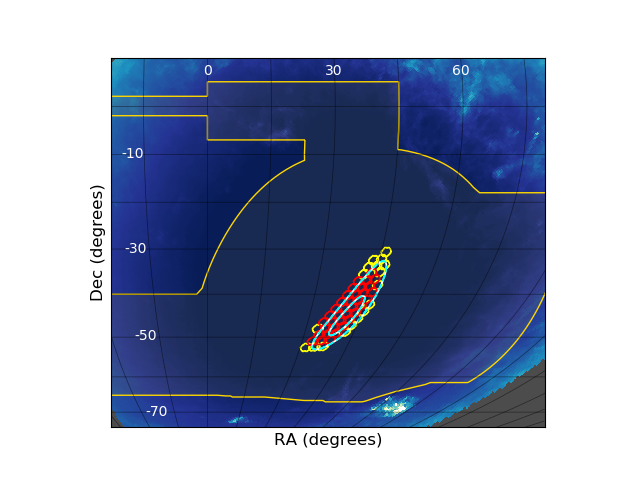}%
\includegraphics[width=0.33\textwidth]{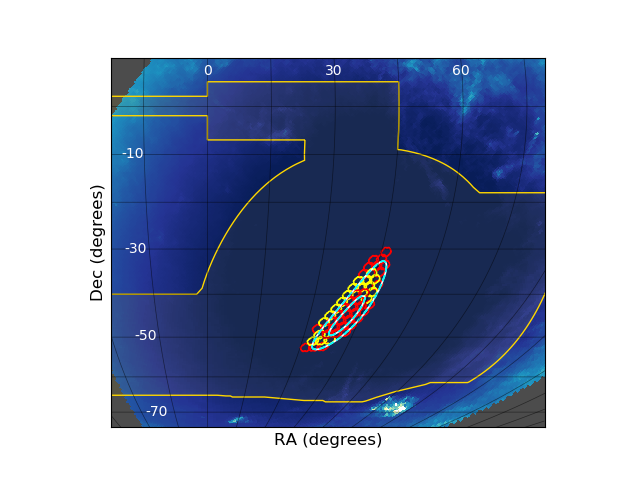}%
\includegraphics[width=0.33\textwidth]{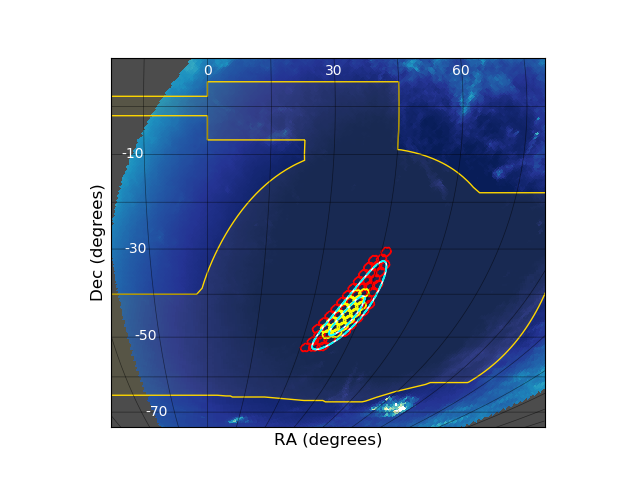}
\caption{Final observing plan for three time slots based on the source detection probability maps in Figs.~\ref{ligoxdes1}, \ref{ligoxdes2}, and \ref{ligoxdes3}). The red hexagons are all scheduled observations across all slots, while the yellow hexagons are those performed within the given time slot. The solid white lines are the 50\%\ and 90\%\ LIGO probability contours, and the yellow line represents the nominal DES footprint. The example shown here is for GW170814, a binary black hole merger.\label{fig:finalobsplan}}
\end{figure*}

To construct the plan we create slots of time that contain integer numbers of hexes
with a total duration of around a half hour; each slot has the full map-making
performed. For NS events there are six hexes per slot (three 90s observations per hex), while for BH and bursts,
there are 18 hexes per slot (one 90s observation per hex). 
The detection probability maps (e.g. Figs.~\ref{ligoxdes1}, \ref{ligoxdes2}, \ref{ligoxdes3}) are then ``hexelated'':
the probabilities are summed inside a fixed pattern of camera pointings.
For the night, the hex with the greatest probability is found and assigned
to be observed in a time slot. The time slot is chosen to maximize the observability of the hex in question (and probability of actual detection of the transient if it happens to be located in those exposures).
That hex is then removed from consideration
(removing all probabilities for that hex at different time slots in the night),
and we repeat the procedure with the next-highest probability hex until all slots are full. We use a mean 
overhead time of 30s between hexes
to account for the time it takes to move the telescope from one position to another.
Now we have a time series that makes full use of observing probabilities,
aiming to ensure we observe as many high-probability hexes as possible 
on a given night, and following an optimal sequence, as shown in Figure~\ref{fig:finalobsplan}.
We convert this list to a JSON file with the appropriate content
to drive DECam and the Blanco telescope. The same JSON files, 
modified for 
a later time, are used during the subsequent observing nights.

\section{Image processing pipeline: images to candidates}\label{sec:proc}

We build on the existing DES image processing pipelines.
For image preparation we use the DES single-epoch (SE) processing 
pipeline~\citep{2018PASP..130g4501M}.
We use a modified version of the DES supernova processing pipeline
(\diffimg{}), described in~\citet{kessler2015}, to perform 
image differencing. This pipeline has a strong 
track record of discovering rare classes of 
transient and rapidly fading 
objects such as in \citet{2017MNRAS.470.4241P} and \citet{2018MNRAS.481..894P}.
We give a description of both pipelines here, focusing on
the modifications made specifically for DESGW.

\subsection{Single-epoch processing}\label{sec:SE}

\begin{figure*}[!htb]
\includegraphics[width=\linewidth]{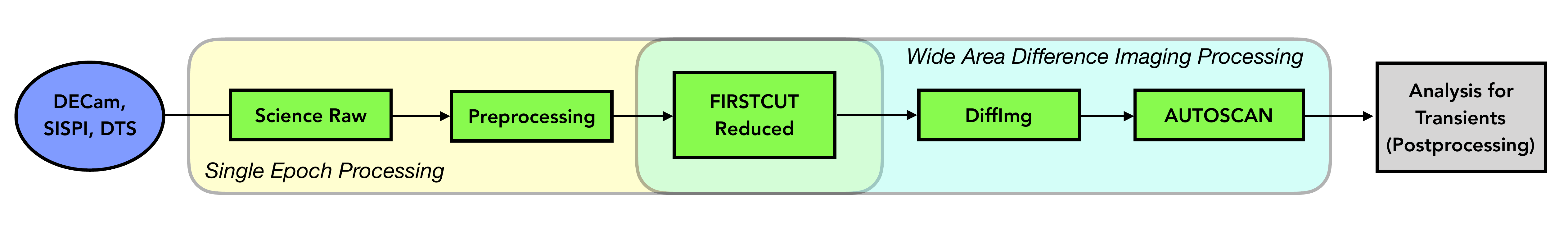}
\caption{
A schematic view of the modules and steps in the DESGW pipeline (shaded area marked ``Wide Area Difference Imaging Processing'').
Calibration and raw science images are delivered from DECam to NCSA. 
In single-epoch processing (yellow box), raw science exposures undergo preprocessing before
proceeding to the \diffimg{} stage (blue box).
Diagram based on Figure 3 of \cite{2018PASP..130g4501M}.
}
\label{fig:single-epoch}
\end{figure*}

The first stage of image processing each night is the SE pipeline, which consists of an image correction stage (\ref{sec:correction}) and an object cataloging stage known as FirstCut (\ref{sec:first-cut}), which includes astrometric calibrations (described separately in~\ref{sec:astrom}).

\subsubsection{Image Correction}\label{sec:correction}
SE begins with a stage to make the raw
images science-ready. This stage includes 
crosstalk corrections, pixel corrections, and bad pixel masking~\citep{Bernstein2017}. The pixel corrections include bias subtraction, 
pixel non-linearity correction, a conversion from DN to electrons, a ``brighter-fatter'' correction, and finally flat fielding. This stage creates a catalog of the brighter objects in each image with {\tt SExtractor}~\citep{2011ASPC..442..435B}, to be used in the astrometric calibration (described in Section \ref{sec:astrom}).

\subsubsection{Astrometric Calibration\label{sec:astrom}}

The astrometry stage uses {\tt SCAMP}~\citep{2006ASPC..351..112B} with the aforementioned catalog of bright objects, generated by {\tt SExtractor} during image correction, to calculate an astrometric solution using an initial-guess of third-order polynomial World Coordinate System (WCS;~\citealt{2002A&A...395.1061G}) distortion terms for each of the 62 CCDs in the DECam array to produce a solution to place CCD pixel positions into a TPV WCS. 
During O1 and O2 we used the 2MASS point source catalog \citep{2006AJ....131.1163S} to solve for the entire focal plane solution for every DES pointing.
We have typical (RMS) DES-2MASS differences of $0.25^{\prime\prime}$. These errors are dominated by 2MASS which 
has typical single detection uncertainties of $0.2^{\prime\prime}$ for sources fainter than 
$K_S$ of 14.
An improvement during the third observing season is to use the Gaia catalog 
\citep{2016A&A...595A...1G,2018A&A...616A...2L} as a reference, allowing both reduction of DES astrometric 
uncertainties to below $0.03^{\prime\prime}$ per coordinate (dominated by DES uncertainties) and for us to calculate an astrometric solution CCD by CCD rather than over the whole image. Per-CCD calculations can run in parallel and are much faster.

\subsubsection{FirstCut}\label{sec:first-cut}
The FirstCut processing calculates an astrometric solution with {\tt SCAMP} (see \ref{sec:astrom}), performs
bleed trail masking, fits and subtracts the sky background, divides
out the star flat, masks cosmic rays and satellite trails, measures and models the
point spread function (PSF), performs object detection and measurement using {\tt SExtractor},
and performs image quality measurements.

To catalog all objects from single-epoch images we run {\tt SExtractor} using PSF modeling and
model-fitting photometry. A PSF model is derived for each CCD image using the {\tt PSFEx}
package~\citep{2011ASPC..442..435B}.  We model PSF variations within each CCD as a $N^\mathrm{th}$ degree
polynomial expansion in CCD coordinates.  For our application we adopt a $26\times26$ pixel kernel and
follow variations to 3$^\mathrm{rd}$ order.

In {\tt SExtractor} (version 2.14.2) we use this PSF model to carry out PSF corrected model
fitting photometry over each image.  The code proceeds by fitting a PSF model and a galaxy model to every
source in the image.  The two-dimensional modeling uses a weighted $\chi^2$ that captures the goodness of
fit between the observed flux distribution and the model and iterates to minimize the $\chi^2$.  The resulting model
parameters are stored and ``asymptotic'' magnitude estimates are extracted by integrating the model flux. 

The advantages of model fitting photometry on single-epoch images that have not
been remapped are manifold.  First, pixel to pixel noise correlations are not 
present in the data and do not have to be corrected for in estimating measurement 
uncertainties.  Second, unbiased PSF and galaxy model fitting photometry is 
available across the image, allowing one to make a more precise 
correction to aperture magnitudes than those often used to extract galaxy and stellar 
photometry.

\subsection{Image differencing}\label{sec:diffimg}

\begin{figure*}[!htbp]
\includegraphics[clip,width=\linewidth]{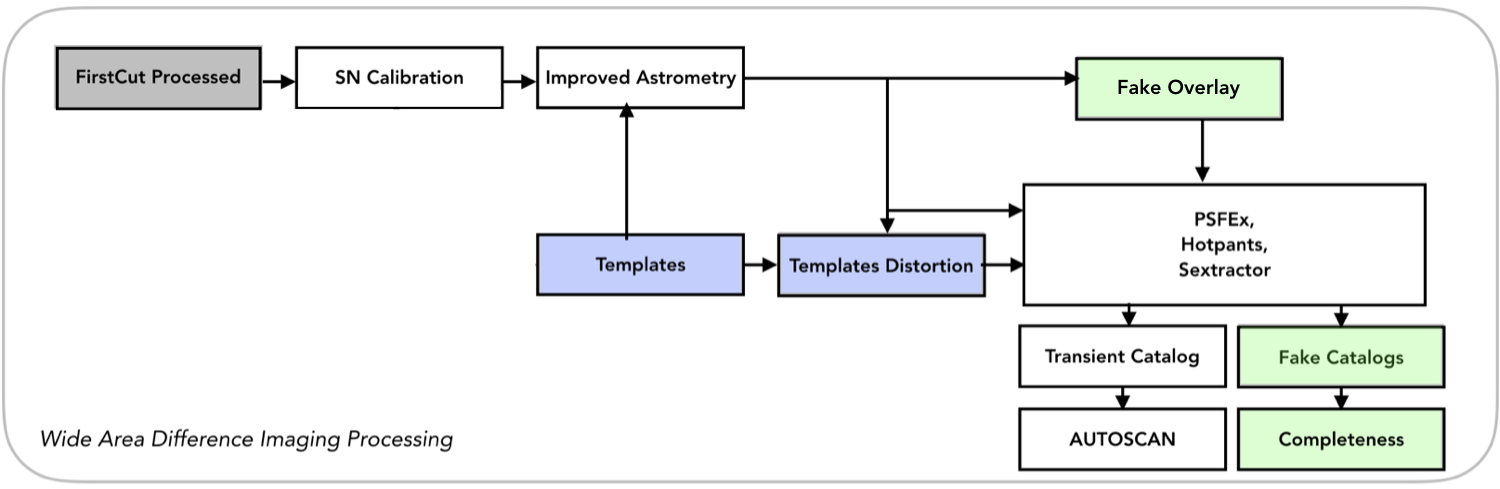}
\caption{
A diagram of the \diffimg{} processing used to detect transients
in the wide area survey and in GW followup fields. Diagram based on Figure 7 of \cite{2018PASP..130g4501M}.
}
\label{diffimg-flow}
\end{figure*}

To identify GW EM counterpart candidates within\\
search images we use the aforementioned \diffimg{}
software, originally developed for supernova searches in DES. Figure~\ref{diffimg-flow} illustrates the workflow. Using a feature of {\tt SCAMP}, we calculate a joint astrometric solution on both the search and template image(s) using the 2MASS (O1 and O2 seasons) or GAIA-DR2 (O3) catalog. We perform the image subtraction via the {\tt HOTPANTS} package \citep{2015ascl.soft04004B}. In our case, we perform a separate subtraction
between the search image and each template, and then combine the difference images, rather than do a single subtraction on one combined template. While this approach is
clearly slower than doing a single subtraction, it avoids potentially large PSF variations that could arise when combining templates taken in potentially very different observing conditions.

\subsubsection{Template preparation}
The \diffimg{} software, also sometimes known as the \diffimg{} ``pipeline", can accept as templates images that only partially 
overlap with search images, images that may have a relative rotation with respect to the search images, and images from DECam that were not
taken on DES time (we only use such images if they have been publicly released). 
These are our main modifications of the pipeline relative to the DES supernova use 
case, where the search and template images are always exactly aligned (within telescope pointing errors) with one of a small set of fixed pointings that comprise the supernova survey area.
After obtaining template images we apply the SE process to them as described in \ref{sec:SE}, so they are ready in the case of a LIGO event trigger. If the counterpart lies outside the DES footprint and no overlapping template image exists, template images of the appropriate area of the sky must be taken at a later time, after we expect any counterpart to have faded.

\subsubsection{Fake injection}
During the \diffimg{} run we inject fake point sources (generated ahead of time) into the search image in each band in random locations away from masked regions. These fakes can either vary in magnitude with time, or be fixed (typically at 20 in that case). Fakes are used for four purposes: to measure the ``completeness depth'', or the depth at which we recover a specified fraction (typically 80\%) of fakes in all of our search images (completeness depth thus reflects the variation in observing conditions); to monitor the single image detection efficiency; to monitor the multiple epoch detection 
efficiency; to characterize \diffimg{} for modeling in a catalog-level MC simulation.

To check our understanding of the efficiency measured with fakes we compare the measured efficiency to a prediction using a simulation from the Supernova Analysis software package, \snana~\citep{2009PASP..121.1028K,2019MNRAS.485.1171K}, 
that analytically computes
light curves without using the detection images, but by using the observed conditions
and key properties of the images derived by measuring the fakes. This allows us to determine how well our sample selection criteria recover objects with known light curves. During O1 
we measured a fake detection completeness of 80\%\ or better down to magnitudes of 22.7 (21.8) in $i$ ($z$) band for 90-second exposures in good observing conditions~\citep{2016ApJ...823L..33S}.

\subsubsection{Candidate identification}\label{sec:candID}
\diffimg{} identifies candidate objects by running\\
{\tt SExtractor} on the difference images.  Objects detected are filtered through a set of selection criteria listed in Table~\ref{table:quality}.
Since the combined difference image includes correlated search image pixels summed over each template, the standard {\tt SExtractor} flux uncertainties are not valid. We developed a special algorithm to properly account for correlations when determining the flux uncertainties.
Surviving objects are referred to as ``detections".
These detections then filter through a machine learning code, \as \citep{2015AJ....150...82G}, which
takes as input the template, search, and difference images and considers such items as
1) the ratio of PSF flux to aperture flux on the template image,
2) the magnitude difference between the detection and the nearest catalog source,
and 3) the {\tt SExtractor}-measured SPREAD\_MODEL of the detection. \as\ returns
a number between 0, an obvious artifact, and 1, a high-quality detection. 

\begin{table*}
\centering
\begin{tabular}{c}
\\
\hline
S/N $>3.5$\\
Object is PSF-like based on SPREAD MODEL\\
SExtractor A\_IMAGE $< 1.5 \times$ PSF\\
In a $35\times35$ box, $<200$ pixels with flux $< -2\sigma$ below $0$\\
In a $35\times35$ box, $<20$ pixels with flux $< -4\sigma$ below $0$\\
In a $35\times35$ box, $<2$ pixels with flux $< -6\sigma$ below $0$\\
Not within a mag-dependent radius of a veto catalog object\\
If coadded, CR rejection via consistency of detected object in each exposure\\
\hline
\end{tabular}
\vspace{-0.2cm} 
\caption{Quality requirements for SExtractor objects detected in search images. From Table 3 of Kessler et al. 2015.}
\label{table:quality}
\end{table*}

\section{Postprocessing}\label{sec:postproc}

\begin{figure*}[!htbp]
\includegraphics[clip,width=\linewidth]{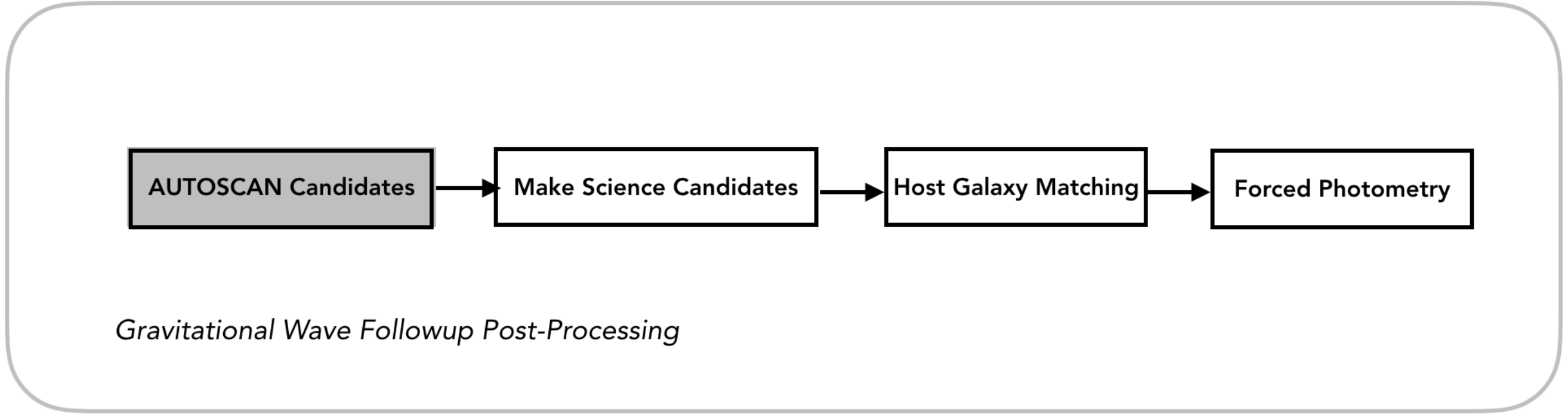}
\caption{
A diagram showing the postprocessing steps. Here the \as\ candidates refer to \diffimg{} detections or ``raw'' candidates, prior to the application of the tighter \as\ requirement. The host galaxy matching runs before forced photometry, but is not required. After forced photometry we can apply additional selection criteria as described in section \ref{sec:postproc}.
}
\label{diffimg-flow2}
\end{figure*}

The outputs of \diffimg{} are the inputs to our post-processing pipeline, illustrated in Fig.~\ref{diffimg-flow2}, which matches detections of the same objects across different exposures and applies quality assurance requirements. It also analyzes fakes injected into the images to assess the performance of \diffimg{}.
This is in preparation for the final sample selection step.

Post-processing takes as input the collection of ``raw" candidates from \diffimg{}, defined as when two or more detections have measured positions matching to within 1 arcsec. The two detections can be in the same band or different bands, or on the same night or different nights. 
All raw candidates are saved, which includes moving objects such as asteroids. Requiring detections on separate nights, or with a minimum time separation on the same night, helps to reject moving objects.
In O1 and O2, we always took images in $i$ and/or $z$ bands in the initial search.

At this point, we also apply a minimum machine learning score requirement, typically 0.7, based on the \texttt{autoScan} score obtained during \diffimg{}. It is a choice to apply the \texttt{autoScan} requirement in post-processing rather than earlier in the process as it facilitates other detection completeness studies with looser requirements. 

``Science candidates'' are those raw candidates that pass the machine learning score requirement. For each science candidate we perform forced PSF photometry at the positions of the science candidates
in all difference images that cover the location. Forced photometry provides flux measurements in all observations, regardless of the S/N.

There exists a ``surface brightness anomaly'' described in both~\cite{kessler2015}
and \cite{doctor2017} which degrades \diffimg{} search efficiency near large galaxies.
The latter reports two effects: 1) excess missed detections of point sources in
areas of high surface brightness as seen in large galaxies and 2) excess flux
scatter in these same areas, which affects the selection of objects in analysis. Strategies to mitigate the effects remain an active area of research.

At this stage the science candidates still contain backgrounds, the leading examples of which are
supernovae, asteroids, and M dwarf flare stars. We now impose
additional selection criteria on the science candidates aimed
at exploiting difference between kilonovae and these backgrounds. Below, we give some example criteria:

\begin{enumerate}
\item Supernovae: supernovae evolve more slowly than kilonovae or plausible BH counterparts. Example criteria to reject them: $\ge 3\sigma$ decline in both $i$ and $z$ fluxes from the first epoch to the second, and $\ge 2\sigma$ flux measurement in both $i$ and $z$ bands in the second epoch. In the case of a single-band search one can make similar flux requirements in the single band, and add additional requirements such as no increase in the candidate's flux in any observation more than 48 hours after the GW event.
\item Asteroids: asteroids move on spatially relevant scales on timescales of minutes. 
Requiring detections separated in time by a sufficient amount, typically 30-60 minutes, eliminates these fast-moving objects. Slow moving asteroids
are rare and faint.
\item M dwarf flare stars: While M stars are red, their flares are very blue, 
essentially 10,000K blackbodies. Example criterion for rejection: $\ge 3\sigma$ decline in $z$-band flux between epochs.
\end{enumerate}
Several other types of objects can mimic a KN signal, as discussed 
in~\cite{2015ApJ...814...25C}, though they generally
have rates two orders of magnitude or lower below the expected SN Ia rates. 

All science candidates undergo a host matching process which identifies nearby 
galaxies and ranks them by probability of being the candidate's host galaxy. This 
matching, implemented only for the latter portion of O2, uses a galaxy catalog 
consisting of data from the SDSS DR13 \citep{2017ApJS..233...25A}, DES Y3Q2 \citep{DES-DR1}, and 2MASS photoz \citep{Bilicki-2013} catalogs.

\section{Grid processing}\label{sec:grid}

In order to meet the turnaround time requirements we must make extensive use
of grid resources. A typical observing night will generate between 100 and 200 new
images, depending on the specifics of the observing plan and weather conditions. 
A typical SE processing job will take around three hours to complete, with \diffimg{}
jobs running on a single CCD taking around one hour. Therefore one night's worth
of images requires between 6,000 and 12,000 CPU hours for the \diffimg{} stage,
and between 300 and 600 additional hours for the SE processing stage. There could be
some additional resources required if there are any template images with incomplete 
SE processing. Our program can effectively use resources at Fermilab
and a wide variety of sites on the Open Science Grid, and can
augment these capabilities by running on commercial clouds if needed. We use a job submission system (Jobsub;~\citealt{jobsub}) that sends GlideinWMS
pilot jobs~\citep{glideinwms} to OSG sites. These pilots are shared among many Fermilab experiments, so due to the combined workload pressure from several experiments rather than just one, there are essentially always pilot jobs running that our workflow can use. We thus largely avoid a long ramp-up time often seen in pilot-based job submissions from a single experiment.

 For a typical observing cadence we expect a new image to be available for processing every three to four minutes. As soon as each new image comes in during followup observations, we run a script that calculates which DECam images overlap with the new image and can be used as templates for \diffimg{} processing, checks whether SE
 processing is complete for that image, and then prepares a multi-stage HTCondor Directed Acyclic Graph that includes SE processing for new 
 image and any incomplete templates in the first stage, with the \diffimg{} processing of the search image in the second stage. 
 In the first stage there is one SE job per search, and in the \diffimg{} stage there is one
 job per CCD, for a total of 60 per search image, which run in parallel. The script also prepares a small
 custom tarball containing the exposure-specific \diffimg{} pipeline scripts and
 information about what template images overlaps with the search image. The script finishes by copying the tarball to 
 Fermilab dCache~\citep{dcache} for use in the corresponding grid jobs, which are then immediately submitted. We distribute our main software stack to worker nodes via the CernVM File System~\citep{2011JPhCS.331d2003B} mechanism.

 \section{Results from O1 and O2}\label{sec:results}

 During the first two LIGO observing seasons, we searched for an optical counterpart
 to a total of four BBH events, two each in O1 and O2, and one BNS event. During O1
we performed two analyses of GW150914~\citep{2016ApJ...823L..33S,2016ApJ...823L..34A}, and one of GW151226~\citep{2016ApJ...826L..29C}.
During O2 we also followed up GW170104 and GW170814~\citep{2019ApJ...873L..24D},
covering 13.6\%\ and 84\%\ of the final LIGO 90\%\ probability regions, respectively.
We also performed followup observations of GW170817~\citep{2017PhRvL.119p1101A}, the first confirmed BNS event from LIGO that ushered in the multi-messenger
astronomy era~\citep{2017ApJ...848L..12A}. Our analysis~\citep{2017ApJ...848L..16S} resulted in one 
of several independent discoveries an optical counterpart near the galaxy NGC4993 11.4h post-merger in $i$ and $z$ bands.
Our team also performed a GW170817 light curve analysis and made a comparison to KN models in~\cite{2017ApJ...848L..17C}. 
We also studied the GW170817 host galaxy, NGC 4993, and its environment in order to understand more about the formation of the GW source~\citep{2017ApJ...849L..34P}. 
The fact that our pipeline detected the 
counterpart, and based on its initial magnitude could have detected a 
similar KN out to 425 Mpc, gives us confidence that we can detect future KNe past
the expected LIGO sensitivity ranges in O3 and O4.

\section{Future directions}\label{sec:future}

We have used the system described in this paper to follow up LIGO sources
from their observing runs O1 and O2. 
Our infrastructure is also well-suited to several other time-domain programs,
including optical followup of high-energy neutrino events from the IceCube
detector. IceCube can localize a neutrino event with a median angular resolution of $\leq 1.0$~sq.~deg. on the sky, an area contained in a single DECam pointing. This program
also aims to rapidly produce candidates for spectroscopic followup by other 
telescopes and is largely based on the pipeline developed here. The program 
released results in~\cite{2019ApJ...883..125M}.

LIGO \& Virgo continue to upgrade their laser interferometers. The full O3 run is expected to 
produce up to $~8$ merger events including neutron stars and
dozens of binary black hole events. The expected rates are large enough to improve constraints on $H_0$, our primary science goal.
These rates also present new challenges for the system
described here, and we have implemented several pipeline upgrades for O3. 
For example we have expanded the available computing resources 
by running the \diffimg{} pipeline on commercial cloud resources. 
As part of the Fermilab HEPCloud project~\citep{Holzman2017}
we ran \diffimg{} processing on Amazon EC2 resources. All tests passed.
We are also adding the ability to incorporate non-DECam template images
to increase our available sky coverage. Thirdly, we have modified the single-epoch processing steps
so that they run on a CCD-by-CCD basis, allowing a single job per CCD to perform both single-epoch 
processing and image differencing. This change leads to fewer jobs overall and reduce the overhead 
associated with multiple jobs and eliminate the fact that in the O1/O2 system, image differencing
would not proceed on any CCD until SE processing was complete on the entire image. In O3, each CCD can be
proceed completely independently, reducing the time to science candidates by as much as a factor of five.
Since astrometry is more difficult on a single CCD, we are using the GAIA-DR2 catalog~\citep{2018A&A...616A...1G} for astrometric calibration in O3.

\section{Summary}

The results from O1 and O2 demonstrate that our infrastructure can
quickly get on sky following a ToO trigger, rapidly process new images, 
and carry out image differencing analysis in a timely manner. The 
infrastructure is not limited to GW event follow-up, however. It is trivial
to apply the same techniques to a wide variety of astrophysical transient
searches, including searches for Trans-Neptunian Objects, the hypothetical 
Planet Nine, and optical signatures from high-energy neutrino events
detected by the IceCube experiment. 

The DESGW program is a partnership between the Dark Energy Survey and members 
of the astronomy community. It is designed to search for electromagnetic 
counterparts to GW events, such as those expected from the merger of two neutron 
stars. DESGW is sensitive to BNS mergers out to approximately 400 Mpc. 2015 and 
early 2016 saw a successful first observing season with optical followup of two 
LVC triggers.  The DESGW program was the most complete in terms of area covered 
and limiting magnitude of all EM counterpart followup programs that studied
GW150914. The program completed a series of improvements to the computing 
infrastructure for followup observation preparation and to the imaging pipeline
itself between the first two LIGO observing seasons. Season two started in November 2016, and saw followup of an additional two BBH events and an independent discovery of the EM counterpart to the first LIGO BNS merger.
The infrastructure also performs well in a variety 
of time-domain astronomy programs. DESGW has excellent potential for discovering additional EM counterparts to 
future GW events, and is eagerly awaiting additional LIGO-Virgo triggers during the 
third observing season. 

\section*{Acknowledgments}
Funding for the DES Projects has been provided by the DOE and NSF (USA), MEC/MICINN/MINECO (Spain), STFC (UK), HEFCE (UK), NCSA (UIUC), KICP (U. Chicago), CCAPP (Ohio State), 
MIFPA (Texas A\&M), CNPQ, FAPERJ, FINEP (Brazil), DFG (Germany) and the Collaborating Institutions in the Dark Energy Survey.

The Collaborating Institutions are Argonne Lab, UC Santa Cruz, University of Cambridge, CIEMAT-Madrid, University of Chicago, University College London, 
DES-Brazil Consortium, University of Edinburgh, ETH Z{\"u}rich, Fermilab, University of Illinois, ICE (IEEC-CSIC), IFAE Barcelona, Lawrence Berkeley Lab, 
LMU M{\"u}nchen and the associated Excellence Cluster Universe, University of Michigan, NOAO, University of Nottingham, Ohio State University, University of 
Pennsylvania, University of Portsmouth, SLAC National Lab, Stanford University, University of Sussex, Texas A\&M University, and the OzDES Membership Consortium.

Based in part on observations at Cerro Tololo Inter-American Observatory
at NSF's NOIRLab, which is managed by the Association of
Universities for Research in Astronomy (AURA) under a cooperative agreement with the National Science Foundation.

The DES Data Management System is supported by the NSF under Grant Numbers AST-1138766 and AST-1536171. The DES participants from Spanish institutions are partially 
supported by MINECO under grants AYA2015-71825, ESP2015-88861, FPA2015-68048, and Centro de Excelencia SEV-2016-0588, SEV-2016-0597 and MDM-2015-0509. Research leading 
to these results has received funding from the ERC under the EU's 7$^{\rm th}$ Framework Programme including grants ERC 240672, 291329 and 306478.
We acknowledge support from the Australian Research Council Centre of Excellence for All-sky Astrophysics (CAASTRO), through project number CE110001020.

This research uses services or data provided by the NOAO Science Archive. NOAO is operated by the Association of Universities for Research in Astronomy (AURA), Inc. under a cooperative agreement with the National Science Foundation.  This manuscript has been authored by Fermi Research Alliance, LLC under Contract No. DE-AC02-07CH11359 with the U.S. Department of Energy, Office of Science, Office of High Energy Physics.  The U.S. Government retains and the publisher, by accepting the article for publication, acknowledges that the U.S. Government retains a non-exclusive, paid-up, irrevocable, world-wide license to publish or reproduce the published form of this manuscript, or allow others to do so, for U.S. Government purposes. 

\bibliography{jta_notes}

\begin{thebibliography}{75}
\expandafter\ifx\csname natexlab\endcsname\relax\def\natexlab#1{#1}\fi
\providecommand{\url}[1]{\texttt{#1}}
\providecommand{\href}[2]{#2}
\providecommand{\path}[1]{#1}
\providecommand{\DOIprefix}{doi:}
\providecommand{\ArXivprefix}{arXiv:}
\providecommand{\URLprefix}{URL: }
\providecommand{\Pubmedprefix}{pmid:}
\providecommand{\doi}[1]{\href{http://dx.doi.org/#1}{\path{#1}}}
\providecommand{\Pubmed}[1]{\href{pmid:#1}{\path{#1}}}
\providecommand{\bibinfo}[2]{#2}
\ifx\xfnm\relax \def\xfnm[#1]{\unskip,\space#1}\fi
\bibitem[{{Abbott} et~al.(2016a){Abbott}, {Abbott}, {Abbott}
  et~al.}]{2016PhRvL.116x1103A}
\bibinfo{author}{{Abbott}, B.P.}, \bibinfo{author}{{Abbott}, R.},
  \bibinfo{author}{{Abbott}, T.D.}, et~al., \bibinfo{year}{2016}a.
\newblock \bibinfo{title}{{GW151226: Observation of Gravitational Waves from a
  22-Solar-Mass Binary Black Hole Coalescence}}.
\newblock \bibinfo{journal}{\prl} \bibinfo{volume}{116},
  \bibinfo{pages}{241103}.
\newblock \DOIprefix\doi{10.1103/PhysRevLett.116.241103},
  \href{http://arxiv.org/abs/1606.04855}{{\tt arXiv:1606.04855}}.
\bibitem[{{Abbott} et~al.(2016b){Abbott}, {Abbott}, {Abbott}
  et~al.}]{2016PhRvL.116f1102A}
\bibinfo{author}{{Abbott}, B.P.}, \bibinfo{author}{{Abbott}, R.},
  \bibinfo{author}{{Abbott}, T.D.}, et~al., \bibinfo{year}{2016}b.
\newblock \bibinfo{title}{{Observation of Gravitational Waves from a Binary
  Black Hole Merger}}.
\newblock \bibinfo{journal}{\prl} \bibinfo{volume}{116},
  \bibinfo{pages}{061102}.
\newblock \DOIprefix\doi{10.1103/PhysRevLett.116.061102},
  \href{http://arxiv.org/abs/1602.03837}{{\tt arXiv:1602.03837}}.
\bibitem[{{Abbott} et~al.(2017a){Abbott}, {Abbott}, {Abbott}
  et~al.}]{2017PhRvL.119p1101A}
\bibinfo{author}{{Abbott}, B.P.}, \bibinfo{author}{{Abbott}, R.},
  \bibinfo{author}{{Abbott}, T.D.}, et~al., \bibinfo{year}{2017}a.
\newblock \bibinfo{title}{{GW170817: Observation of Gravitational Waves from a
  Binary Neutron Star Inspiral}}.
\newblock \bibinfo{journal}{Phys.~Rev.~Lett.} \bibinfo{volume}{119},
  \bibinfo{pages}{161101}.
\newblock \DOIprefix\doi{10.1103/PhysRevLett.119.161101},
  \href{http://arxiv.org/abs/1710.05832}{{\tt arXiv:1710.05832}}.
\bibitem[{{Abbott} et~al.(2017b){Abbott}, {Abbott}
  et~al.}]{2017ApJ...848L..12A}
\bibinfo{author}{{Abbott}, B.P.}, \bibinfo{author}{{Abbott}, R.}, et~al.,
  \bibinfo{year}{2017}b.
\newblock \bibinfo{title}{{Multi-messenger Observations of a Binary Neutron
  Star Merger}}.
\newblock \bibinfo{journal}{\apj} \bibinfo{volume}{848}, \bibinfo{pages}{L12}.
\newblock \DOIprefix\doi{10.3847/2041-8213/aa91c9},
  \href{http://arxiv.org/abs/1710.05833}{{\tt arXiv:1710.05833}}.
\bibitem[{{Abbott} et~al.(2018){Abbott}, {Abdalla}, {Allam} et~al.}]{DES-DR1}
\bibinfo{author}{{Abbott}, T.M.C.}, \bibinfo{author}{{Abdalla}, F.B.},
  \bibinfo{author}{{Allam}, S.}, et~al., \bibinfo{year}{2018}.
\newblock \bibinfo{title}{The dark energy survey: Data release 1}.
\newblock \bibinfo{journal}{\apjs} \bibinfo{volume}{239}, \bibinfo{pages}{18}.
\newblock \DOIprefix\doi{10.3847/1538-4365/aae9f0}.
\bibitem[{{Albareti} et~al.(2017){Albareti}, {Allende Prieto}, {Almeida}
  et~al.}]{2017ApJS..233...25A}
\bibinfo{author}{{Albareti}, F.D.}, \bibinfo{author}{{Allende Prieto}, C.},
  \bibinfo{author}{{Almeida}, A.}, et~al., \bibinfo{year}{2017}.
\newblock \bibinfo{title}{{The 13th Data Release of the Sloan Digital Sky
  Survey: First Spectroscopic Data from the SDSS-IV Survey Mapping Nearby
  Galaxies at Apache Point Observatory}}.
\newblock \bibinfo{journal}{ApJs} \bibinfo{volume}{233}, \bibinfo{pages}{25}.
\newblock \DOIprefix\doi{10.3847/1538-4365/aa8992},
  \href{http://arxiv.org/abs/1608.02013}{{\tt arXiv:1608.02013}}.
\bibitem[{{Annis} et~al.(2016){Annis}, {Soares-Santos}, {Berger}
  et~al.}]{2016ApJ...823L..34A}
\bibinfo{author}{{Annis}, J.}, \bibinfo{author}{{Soares-Santos}, M.},
  \bibinfo{author}{{Berger}, E.}, et~al., \bibinfo{year}{2016}.
\newblock \bibinfo{title}{{A Dark Energy Camera Search for Missing Supergiants
  in the LMC after the Advanced LIGO Gravitational-wave Event GW150914}}.
\newblock \bibinfo{journal}{\apj} \bibinfo{volume}{823}, \bibinfo{pages}{L34}.
\newblock \DOIprefix\doi{10.3847/2041-8205/823/2/L34},
  \href{http://arxiv.org/abs/1602.04199}{{\tt arXiv:1602.04199}}.
\bibitem[{{Barnes} and {Kasen}(2013)}]{barnes2013}
\bibinfo{author}{{Barnes}, J.}, \bibinfo{author}{{Kasen}, D.},
  \bibinfo{year}{2013}.
\newblock \bibinfo{title}{{Effect of a High Opacity on the Light Curves of
  Radioactively Powered Transients from Compact Object Mergers}}.
\newblock \bibinfo{journal}{\apj} \bibinfo{volume}{775}, \bibinfo{pages}{18}.
\newblock \DOIprefix\doi{10.1088/0004-637X/775/1/18},
  \href{http://arxiv.org/abs/1303.5787}{{\tt arXiv:1303.5787}}.
\bibitem[{{Barnes} et~al.(2016){Barnes}, {Kasen}, {Wu} and {Mart{\'\i
  }nez-Pinedo}}]{2016ApJ...829..110B}
\bibinfo{author}{{Barnes}, J.}, \bibinfo{author}{{Kasen}, D.},
  \bibinfo{author}{{Wu}, M.R.}, \bibinfo{author}{{Mart{\'\i }nez-Pinedo}, G.},
  \bibinfo{year}{2016}.
\newblock \bibinfo{title}{{Radioactivity and Thermalization in the Ejecta of
  Compact Object Mergers and Their Impact on Kilonova Light Curves}}.
\newblock \bibinfo{journal}{\apj} \bibinfo{volume}{829}, \bibinfo{pages}{110}.
\newblock \DOIprefix\doi{10.3847/0004-637X/829/2/110},
  \href{http://arxiv.org/abs/1605.07218}{{\tt arXiv:1605.07218}}.
\bibitem[{{Becker}(2015)}]{2015ascl.soft04004B}
\bibinfo{author}{{Becker}, A.}, \bibinfo{year}{2015}.
\newblock \bibinfo{title}{{HOTPANTS: High Order Transform of PSF ANd Template
  Subtraction}}.
\newblock \bibinfo{howpublished}{Astrophysics Source Code Library}.
\newblock \URLprefix \url{https://github.com/acbecker/hotpants},
  \href{http://arxiv.org/abs/1504.004}{{\tt arXiv:1504.004}}.
\bibitem[{{Berger}(2014)}]{2014ARA&A..52...43B}
\bibinfo{author}{{Berger}, E.}, \bibinfo{year}{2014}.
\newblock \bibinfo{title}{{Short-Duration Gamma-Ray Bursts}}.
\newblock \bibinfo{journal}{Annu. Rev. Astron. Astrophys.}
  \bibinfo{volume}{52}, \bibinfo{pages}{43--105}.
\newblock \DOIprefix\doi{10.1146/annurev-astro-081913-035926},
  \href{http://arxiv.org/abs/1311.2603}{{\tt arXiv:1311.2603}}.
\bibitem[{{Bernstein} et~al.(2018){Bernstein}, {Abbott}, {Armstrong}
  et~al.}]{Bernstein2017}
\bibinfo{author}{{Bernstein}, G.M.}, \bibinfo{author}{{Abbott}, T.M.C.},
  \bibinfo{author}{{Armstrong}, R.}, et~al., \bibinfo{year}{2018}.
\newblock \bibinfo{title}{{Photometric Characterization of the Dark Energy
  Camera}}.
\newblock \bibinfo{journal}{\pasp} \bibinfo{volume}{130},
  \bibinfo{pages}{054501}.
\newblock \DOIprefix\doi{10.1088/1538-3873/aaa753},
  \href{http://arxiv.org/abs/1710.10943}{{\tt arXiv:1710.10943}}.
\bibitem[{{Bertin}(2006)}]{2006ASPC..351..112B}
\bibinfo{author}{{Bertin}, E.}, \bibinfo{year}{2006}.
\newblock \bibinfo{title}{{Automatic Astrometric and Photometric Calibration
  with SCAMP}}. volume \bibinfo{volume}{351} of
  \textit{\bibinfo{series}{Astron. Soc. Pac. Conf. Ser.}}
\newblock p. \bibinfo{pages}{112}.
\bibitem[{{Bertin}(2011)}]{2011ASPC..442..435B}
\bibinfo{author}{{Bertin}, E.}, \bibinfo{year}{2011}.
\newblock \bibinfo{title}{{Automated Morphometry with SExtractor and PSFEx}},
  in: \bibinfo{editor}{{Evans}, I.N.}, \bibinfo{editor}{{Accomazzi}, A.},
  \bibinfo{editor}{{Mink}, D.J.}, \bibinfo{editor}{{Rots}, A.H.} (Eds.),
  \bibinfo{booktitle}{Astronomical Data Analysis Software and Systems XX}, p.
  \bibinfo{pages}{435}.
\bibitem[{{Bilicki} et~al.(2014){Bilicki}, {Jarrett}, {Peacock}
  et~al.}]{Bilicki-2013}
\bibinfo{author}{{Bilicki}, M.}, \bibinfo{author}{{Jarrett}, T.H.},
  \bibinfo{author}{{Peacock}, J.A.}, et~al., \bibinfo{year}{2014}.
\newblock \bibinfo{title}{{Two Micron All Sky Survey Photometric Redshift
  Catalog: A Comprehensive Three-dimensional Census of the Whole Sky}}.
\newblock \bibinfo{journal}{\apjs} \bibinfo{volume}{210}, \bibinfo{pages}{9}.
\newblock \DOIprefix\doi{10.1088/0067-0049/210/1/9},
  \href{http://arxiv.org/abs/1311.5246}{{\tt arXiv:1311.5246}}.
\bibitem[{{Blomer} et~al.(2011){Blomer}, {Aguado-S{\'a}nchez}, {Buncic} and
  {Harutyunyan}}]{2011JPhCS.331d2003B}
\bibinfo{author}{{Blomer}, J.}, \bibinfo{author}{{Aguado-S{\'a}nchez}, C.},
  \bibinfo{author}{{Buncic}, P.}, \bibinfo{author}{{Harutyunyan}, A.},
  \bibinfo{year}{2011}.
\newblock \bibinfo{title}{{Distributing LHC application software and conditions
  databases using the CernVM file system}}, in: \bibinfo{booktitle}{J. Phys.:
  Conf. Ser.}, p. \bibinfo{pages}{042003}.
\newblock \DOIprefix\doi{10.1088/1742-6596/331/4/042003}.
\bibitem[{{Box}(2014)}]{jobsub}
\bibinfo{author}{{Box}, D.}, \bibinfo{year}{2014}.
\newblock \bibinfo{title}{{FIFE-Jobsub: a grid submission system for intensity
  frontier experiments at Fermilab}}.
\newblock \bibinfo{journal}{J. Phys.: Conf. Ser.} \bibinfo{volume}{513},
  \bibinfo{pages}{032010}.
\newblock \DOIprefix\doi{10.1088/1742-6596/513/3/032010}.
\bibitem[{{Bulla}(2019)}]{2019arXiv190604205B}
\bibinfo{author}{{Bulla}, M.}, \bibinfo{year}{2019}.
\newblock \bibinfo{title}{{POSSIS: predicting spectra, light curves and
  polarization for multi-dimensional models of supernovae and kilonovae}}.
\newblock \bibinfo{journal}{arXiv e-prints} ,
  \bibinfo{pages}{arXiv:1906.04205}\href{http://arxiv.org/abs/1906.04205}{{\tt
  arXiv:1906.04205}}.
\bibitem[{{Chen} et~al.(2018){Chen}, {Fishbach} and
  {Holz}}]{2018Natur.562..545C}
\bibinfo{author}{{Chen}, H.Y.}, \bibinfo{author}{{Fishbach}, M.},
  \bibinfo{author}{{Holz}, D.E.}, \bibinfo{year}{2018}.
\newblock \bibinfo{title}{{A two per cent Hubble constant measurement from
  standard sirens within five years}}.
\newblock \bibinfo{journal}{Nature} \bibinfo{volume}{562},
  \bibinfo{pages}{545--547}.
\newblock \DOIprefix\doi{10.1038/s41586-018-0606-0},
  \href{http://arxiv.org/abs/1712.06531}{{\tt arXiv:1712.06531}}.
\bibitem[{{Cowperthwaite} and {Berger}(2015)}]{2015ApJ...814...25C}
\bibinfo{author}{{Cowperthwaite}, P.S.}, \bibinfo{author}{{Berger}, E.},
  \bibinfo{year}{2015}.
\newblock \bibinfo{title}{{A Comprehensive Study of Detectability and
  Contamination in Deep Rapid Optical Searches for Gravitational Wave
  Counterparts}}.
\newblock \bibinfo{journal}{\apj} \bibinfo{volume}{814}, \bibinfo{pages}{25}.
\newblock \DOIprefix\doi{10.1088/0004-637X/814/1/25},
  \href{http://arxiv.org/abs/1503.07869}{{\tt arXiv:1503.07869}}.
\bibitem[{{Cowperthwaite} et~al.(2016){Cowperthwaite}, {Berger},
  {Soares-Santos} et~al.}]{2016ApJ...826L..29C}
\bibinfo{author}{{Cowperthwaite}, P.S.}, \bibinfo{author}{{Berger}, E.},
  \bibinfo{author}{{Soares-Santos}, M.}, et~al., \bibinfo{year}{2016}.
\newblock \bibinfo{title}{{A DECam Search for an Optical Counterpart to the
  LIGO Gravitational-wave Event GW151226}}.
\newblock \bibinfo{journal}{\apj} \bibinfo{volume}{826}, \bibinfo{pages}{L29}.
\newblock \DOIprefix\doi{10.3847/2041-8205/826/2/L29},
  \href{http://arxiv.org/abs/1606.04538}{{\tt arXiv:1606.04538}}.
\bibitem[{{Cowperthwaite} et~al.(2017){Cowperthwaite}, {Berger}, {Villar}
  et~al.}]{2017ApJ...848L..17C}
\bibinfo{author}{{Cowperthwaite}, P.S.}, \bibinfo{author}{{Berger}, E.},
  \bibinfo{author}{{Villar}, V.A.}, et~al., \bibinfo{year}{2017}.
\newblock \bibinfo{title}{{The Electromagnetic Counterpart of the Binary
  Neutron Star Merger LIGO/Virgo GW170817. II. UV, Optical, and Near-infrared
  Light Curves and Comparison to Kilonova Models}}.
\newblock \bibinfo{journal}{\apj} \bibinfo{volume}{848}, \bibinfo{pages}{L17}.
\newblock \DOIprefix\doi{10.3847/2041-8213/aa8fc7},
  \href{http://arxiv.org/abs/1710.05840}{{\tt arXiv:1710.05840}}.
\bibitem[{{Del Pozzo}(2012)}]{2012PhRvD..86d3011D}
\bibinfo{author}{{Del Pozzo}, W.}, \bibinfo{year}{2012}.
\newblock \bibinfo{title}{{Inference of cosmological parameters from
  gravitational waves: Applications to second generation interferometers}}.
\newblock \bibinfo{journal}{\prd} \bibinfo{volume}{86},
  \bibinfo{pages}{043011}.
\newblock \DOIprefix\doi{10.1103/PhysRevD.86.043011},
  \href{http://arxiv.org/abs/1108.1317}{{\tt arXiv:1108.1317}}.
\bibitem[{{Doctor} et~al.(2017){Doctor}, {Kessler}, {Chen} et~al.}]{doctor2017}
\bibinfo{author}{{Doctor}, Z.}, \bibinfo{author}{{Kessler}, R.},
  \bibinfo{author}{{Chen}, H.Y.}, et~al., \bibinfo{year}{2017}.
\newblock \bibinfo{title}{{A Search for Kilonovae in the Dark Energy Survey}}.
\newblock \bibinfo{journal}{\apj} \bibinfo{volume}{837}, \bibinfo{pages}{57}.
\newblock \DOIprefix\doi{10.3847/1538-4357/aa5d09},
  \href{http://arxiv.org/abs/1611.08052}{{\tt arXiv:1611.08052}}.
\bibitem[{{Doctor} et~al.(2019){Doctor}, {Kessler}, {Herner}
  et~al.}]{2019ApJ...873L..24D}
\bibinfo{author}{{Doctor}, Z.}, \bibinfo{author}{{Kessler}, R.},
  \bibinfo{author}{{Herner}, K.}, et~al., \bibinfo{year}{2019}.
\newblock \bibinfo{title}{{A Search for Optical Emission from Binary Black Hole
  Merger GW170814 with the Dark Energy Camera}}.
\newblock \bibinfo{journal}{\apjl} \bibinfo{volume}{873}, \bibinfo{pages}{L24}.
\newblock \DOIprefix\doi{10.3847/2041-8213/ab08a3},
  \href{http://arxiv.org/abs/1812.01579}{{\tt arXiv:1812.01579}}.
\bibitem[{{Duque} et~al.(2019){Duque}, {Daigne} and
  {Mochkovitch}}]{2019arXiv190504495D}
\bibinfo{author}{{Duque}, R.}, \bibinfo{author}{{Daigne}, F.},
  \bibinfo{author}{{Mochkovitch}, R.}, \bibinfo{year}{2019}.
\newblock \bibinfo{title}{{Predictions for radio afterglows of binary neutron
  star mergers: a population study for O3 and beyond}}.
\newblock \bibinfo{journal}{arXiv e-prints} ,
  \bibinfo{pages}{arXiv:1905.04495}\href{http://arxiv.org/abs/1905.04495}{{\tt
  arXiv:1905.04495}}.
\bibitem[{{Fahlman} and {Fern{\'a}ndez}(2018)}]{2018ApJ...869L...3F}
\bibinfo{author}{{Fahlman}, S.}, \bibinfo{author}{{Fern{\'a}ndez}, R.},
  \bibinfo{year}{2018}.
\newblock \bibinfo{title}{{Hypermassive Neutron Star Disk Outflows and Blue
  Kilonovae}}.
\newblock \bibinfo{journal}{\apjl} \bibinfo{volume}{869}, \bibinfo{pages}{L3}.
\newblock \DOIprefix\doi{10.3847/2041-8213/aaf1ab},
  \href{http://arxiv.org/abs/1811.08906}{{\tt arXiv:1811.08906}}.
\bibitem[{{Flaugher} et~al.(2015){Flaugher}, {Diehl}, {Honscheid}
  et~al.}]{flaugher2015}
\bibinfo{author}{{Flaugher}, B.}, \bibinfo{author}{{Diehl}, H.T.},
  \bibinfo{author}{{Honscheid}, K.}, et~al., \bibinfo{year}{2015}.
\newblock \bibinfo{title}{{The Dark Energy Camera}}.
\newblock \bibinfo{journal}{\aj} \bibinfo{volume}{150}, \bibinfo{pages}{150}.
\newblock \DOIprefix\doi{10.1088/0004-6256/150/5/150},
  \href{http://arxiv.org/abs/1504.02900}{{\tt arXiv:1504.02900}}.
\bibitem[{Fuhrmann and G{\"u}lzow(2006)}]{dcache}
\bibinfo{author}{Fuhrmann, P.}, \bibinfo{author}{G{\"u}lzow, V.},
  \bibinfo{year}{2006}.
\newblock \bibinfo{title}{dcache, storage system for the future}, in:
  \bibinfo{editor}{Nagel, W.E.}, \bibinfo{editor}{Walter, W.V.},
  \bibinfo{editor}{Lehner, W.} (Eds.), \bibinfo{booktitle}{Euro-Par 2006
  Parallel Processing}, \bibinfo{publisher}{Springer Berlin Heidelberg},
  \bibinfo{address}{Berlin, Heidelberg}. pp. \bibinfo{pages}{1106--1113}.
\bibitem[{{Gaertig} et~al.(2011){Gaertig}, {Glampedakis}, {Kokkotas} and
  {Zink}}]{2011PhRvL.107j1102G}
\bibinfo{author}{{Gaertig}, E.}, \bibinfo{author}{{Glampedakis}, K.},
  \bibinfo{author}{{Kokkotas}, K.D.}, \bibinfo{author}{{Zink}, B.},
  \bibinfo{year}{2011}.
\newblock \bibinfo{title}{{f-Mode Instability in Relativistic Neutron Stars}}.
\newblock \bibinfo{journal}{\prl} \bibinfo{volume}{107},
  \bibinfo{pages}{101102}.
\newblock \DOIprefix\doi{10.1103/PhysRevLett.107.101102},
  \href{http://arxiv.org/abs/1106.5512}{{\tt arXiv:1106.5512}}.
\bibitem[{{Gaia Collaboration} et~al.(2018){Gaia Collaboration}, {Brown},
  {Vallenari}, {Prusti} et~al.}]{2018A&A...616A...1G}
\bibinfo{author}{{Gaia Collaboration}}, \bibinfo{author}{{Brown}, A.G.A.},
  \bibinfo{author}{{Vallenari}, A.}, \bibinfo{author}{{Prusti}, T.}, et~al.,
  \bibinfo{year}{2018}.
\newblock \bibinfo{title}{{Gaia Data Release 2. Summary of the contents and
  survey properties}}.
\newblock \bibinfo{journal}{\aap} \bibinfo{volume}{616}, \bibinfo{pages}{A1}.
\newblock \DOIprefix\doi{10.1051/0004-6361/201833051},
  \href{http://arxiv.org/abs/1804.09365}{{\tt arXiv:1804.09365}}.
\bibitem[{{Gaia Collaboration} et~al.(2016){Gaia Collaboration}, {Prusti}, {de
  Bruijne}, {Brown} et~al.}]{2016A&A...595A...1G}
\bibinfo{author}{{Gaia Collaboration}}, \bibinfo{author}{{Prusti}, T.},
  \bibinfo{author}{{de Bruijne}, J.H.J.}, \bibinfo{author}{{Brown}, A.G.A.},
  et~al., \bibinfo{year}{2016}.
\newblock \bibinfo{title}{{The Gaia mission}}.
\newblock \bibinfo{journal}{\aap} \bibinfo{volume}{595}, \bibinfo{pages}{A1}.
\newblock \DOIprefix\doi{10.1051/0004-6361/201629272},
  \href{http://arxiv.org/abs/1609.04153}{{\tt arXiv:1609.04153}}.
\bibitem[{{Goldstein} et~al.(2015){Goldstein}, {D'Andrea}, {Fischer}
  et~al.}]{2015AJ....150...82G}
\bibinfo{author}{{Goldstein}, D.A.}, \bibinfo{author}{{D'Andrea}, C.B.},
  \bibinfo{author}{{Fischer}, J.A.}, et~al., \bibinfo{year}{2015}.
\newblock \bibinfo{title}{{Automated Transient Identification in the Dark
  Energy Survey}}.
\newblock \bibinfo{journal}{\aj} \bibinfo{volume}{150}, \bibinfo{pages}{82}.
\newblock \DOIprefix\doi{10.1088/0004-6256/150/3/82},
  \href{http://arxiv.org/abs/1504.02936}{{\tt arXiv:1504.02936}}.
\bibitem[{{Gompertz} et~al.(2018){Gompertz}, {Levan}, {Tanvir}
  et~al.}]{2018ApJ...860...62G}
\bibinfo{author}{{Gompertz}, B.P.}, \bibinfo{author}{{Levan}, A.J.},
  \bibinfo{author}{{Tanvir}, N.R.}, et~al., \bibinfo{year}{2018}.
\newblock \bibinfo{title}{{The Diversity of Kilonova Emission in Short
  Gamma-Ray Bursts}}.
\newblock \bibinfo{journal}{\apj} \bibinfo{volume}{860}, \bibinfo{pages}{62}.
\newblock \DOIprefix\doi{10.3847/1538-4357/aac206},
  \href{http://arxiv.org/abs/1710.05442}{{\tt arXiv:1710.05442}}.
\bibitem[{{G{\'o}rski} et~al.(2005){G{\'o}rski}, {Hivon}, {Banday}
  et~al.}]{2005ApJ...622..759G}
\bibinfo{author}{{G{\'o}rski}, K.M.}, \bibinfo{author}{{Hivon}, E.},
  \bibinfo{author}{{Banday}, A.J.}, et~al., \bibinfo{year}{2005}.
\newblock \bibinfo{title}{{HEALPix: A Framework for High-Resolution
  Discretization and Fast Analysis of Data Distributed on the Sphere}}.
\newblock \bibinfo{journal}{\apj} \bibinfo{volume}{622},
  \bibinfo{pages}{759--771}.
\newblock \DOIprefix\doi{10.1086/427976},
  \href{http://arxiv.org/abs/astro-ph/0409513}{{\tt arXiv:astro-ph/0409513}}.
\bibitem[{{Gottlieb} et~al.(2018){Gottlieb}, {Nakar} and
  {Piran}}]{2018MNRAS.473..576G}
\bibinfo{author}{{Gottlieb}, O.}, \bibinfo{author}{{Nakar}, E.},
  \bibinfo{author}{{Piran}, T.}, \bibinfo{year}{2018}.
\newblock \bibinfo{title}{{The cocoon emission - an electromagnetic counterpart
  to gravitational waves from neutron star mergers}}.
\newblock \bibinfo{journal}{\mnras} \bibinfo{volume}{473},
  \bibinfo{pages}{576--584}.
\newblock \DOIprefix\doi{10.1093/mnras/stx2357},
  \href{http://arxiv.org/abs/1705.10797}{{\tt arXiv:1705.10797}}.
\bibitem[{{Greisen} and {Calabretta}(2002)}]{2002A&A...395.1061G}
\bibinfo{author}{{Greisen}, E.W.}, \bibinfo{author}{{Calabretta}, M.R.},
  \bibinfo{year}{2002}.
\newblock \bibinfo{title}{{Representations of world coordinates in FITS}}.
\newblock \bibinfo{journal}{\aap} \bibinfo{volume}{395},
  \bibinfo{pages}{1061--1075}.
\newblock \DOIprefix\doi{10.1051/0004-6361:20021326},
  \href{http://arxiv.org/abs/astro-ph/0207407}{{\tt arXiv:astro-ph/0207407}}.
\bibitem[{{Grossman} et~al.(2014){Grossman}, {Korobkin}, {Rosswog} and
  {Piran}}]{grossman2014}
\bibinfo{author}{{Grossman}, D.}, \bibinfo{author}{{Korobkin}, O.},
  \bibinfo{author}{{Rosswog}, S.}, \bibinfo{author}{{Piran}, T.},
  \bibinfo{year}{2014}.
\newblock \bibinfo{title}{{The long-term evolution of neutron star merger
  remnants - II. Radioactively powered transients}}.
\newblock \bibinfo{journal}{\mnras} \bibinfo{volume}{439},
  \bibinfo{pages}{757--770}.
\newblock \DOIprefix\doi{10.1093/mnras/stt2503},
  \href{http://arxiv.org/abs/1307.2943}{{\tt arXiv:1307.2943}}.
\bibitem[{{Holz} and {Hughes}(2005)}]{Holz2005}
\bibinfo{author}{{Holz}, D.E.}, \bibinfo{author}{{Hughes}, S.A.},
  \bibinfo{year}{2005}.
\newblock \bibinfo{title}{Using gravitational-wave standard sirens}.
\newblock \bibinfo{journal}{\apj} \bibinfo{volume}{629}, \bibinfo{pages}{15}.
\newblock \URLprefix \url{http://stacks.iop.org/0004-637X/629/i=1/a=15}.
\bibitem[{Holzman et~al.(2017)Holzman, Bauerdick, Bockelman
  et~al.}]{Holzman2017}
\bibinfo{author}{Holzman, B.}, \bibinfo{author}{Bauerdick, L.A.T.},
  \bibinfo{author}{Bockelman, B.}, et~al., \bibinfo{year}{2017}.
\newblock \bibinfo{title}{Hepcloud, a new paradigm for {HEP} facilities: {CMS}
  {Amazon Web Services} investigation}.
\newblock \bibinfo{journal}{Comput. and Softw. for Big Sci.}
  \bibinfo{volume}{1}, \bibinfo{pages}{1}.
\newblock \URLprefix \url{https://doi.org/10.1007/s41781-017-0001-9},
  \DOIprefix\doi{10.1007/s41781-017-0001-9}.
\bibitem[{{Kessler} et~al.(2009){Kessler}, {Bernstein}, {Cinabro}
  et~al.}]{2009PASP..121.1028K}
\bibinfo{author}{{Kessler}, R.}, \bibinfo{author}{{Bernstein}, J.P.},
  \bibinfo{author}{{Cinabro}, D.}, et~al., \bibinfo{year}{2009}.
\newblock \bibinfo{title}{{SNANA: A Public Software Package for Supernova
  Analysis}}.
\newblock \bibinfo{journal}{\pasp} \bibinfo{volume}{121},
  \bibinfo{pages}{1028}.
\newblock \DOIprefix\doi{10.1086/605984},
  \href{http://arxiv.org/abs/0908.4280}{{\tt arXiv:0908.4280}}.
\bibitem[{{Kessler} et~al.(2019){Kessler}, {Brout}, {D'Andrea}
  et~al.}]{2019MNRAS.485.1171K}
\bibinfo{author}{{Kessler}, R.}, \bibinfo{author}{{Brout}, D.},
  \bibinfo{author}{{D'Andrea}, C.B.}, et~al., \bibinfo{year}{2019}.
\newblock \bibinfo{title}{{First cosmology results using Type Ia supernova from
  the Dark Energy Survey: simulations to correct supernova distance biases}}.
\newblock \bibinfo{journal}{\mnras} \bibinfo{volume}{485},
  \bibinfo{pages}{1171--1187}.
\newblock \DOIprefix\doi{10.1093/mnras/stz463},
  \href{http://arxiv.org/abs/1811.02379}{{\tt arXiv:1811.02379}}.
\bibitem[{{Kessler} et~al.(2015){Kessler}, {Marriner}, {Childress}
  et~al.}]{kessler2015}
\bibinfo{author}{{Kessler}, R.}, \bibinfo{author}{{Marriner}, J.},
  \bibinfo{author}{{Childress}, M.}, et~al., \bibinfo{year}{2015}.
\newblock \bibinfo{title}{{The Difference Imaging Pipeline for the Transient
  Search in the Dark Energy Survey}}.
\newblock \bibinfo{journal}{\aj} \bibinfo{volume}{150}, \bibinfo{pages}{172}.
\newblock \DOIprefix\doi{10.1088/0004-6256/150/6/172},
  \href{http://arxiv.org/abs/1507.05137}{{\tt arXiv:1507.05137}}.
\bibitem[{{Krisciunas} and {Schaefer}(1991)}]{krisciunas1991}
\bibinfo{author}{{Krisciunas}, K.}, \bibinfo{author}{{Schaefer}, B.E.},
  \bibinfo{year}{1991}.
\newblock \bibinfo{title}{{A model of the brightness of moonlight}}.
\newblock \bibinfo{journal}{\pasp} \bibinfo{volume}{103},
  \bibinfo{pages}{1033--1039}.
\newblock \DOIprefix\doi{10.1086/132921}.
\bibitem[{{Lindegren} et~al.(2018){Lindegren}, {Hern{\'a}ndez}, {Bombrun}
  et~al.}]{2018A&A...616A...2L}
\bibinfo{author}{{Lindegren}, L.}, \bibinfo{author}{{Hern{\'a}ndez}, J.},
  \bibinfo{author}{{Bombrun}, A.}, et~al., \bibinfo{year}{2018}.
\newblock \bibinfo{title}{{Gaia Data Release 2. The astrometric solution}}.
\newblock \bibinfo{journal}{\aap} \bibinfo{volume}{616}, \bibinfo{pages}{A2}.
\newblock \DOIprefix\doi{10.1051/0004-6361/201832727},
  \href{http://arxiv.org/abs/1804.09366}{{\tt arXiv:1804.09366}}.
\bibitem[{{MacLeod} and {Hogan}(2008)}]{2008PhRvD..77d3512M}
\bibinfo{author}{{MacLeod}, C.L.}, \bibinfo{author}{{Hogan}, C.J.},
  \bibinfo{year}{2008}.
\newblock \bibinfo{title}{{Precision of Hubble constant derived using black
  hole binary absolute distances and statistical redshift information}}.
\newblock \bibinfo{journal}{\prd} \bibinfo{volume}{77},
  \bibinfo{pages}{043512}.
\newblock \DOIprefix\doi{10.1103/PhysRevD.77.043512},
  \href{http://arxiv.org/abs/0712.0618}{{\tt arXiv:0712.0618}}.
\bibitem[{{Metzger}(2017)}]{2017LRR....20....3M}
\bibinfo{author}{{Metzger}, B.D.}, \bibinfo{year}{2017}.
\newblock \bibinfo{title}{{Kilonovae}}.
\newblock \bibinfo{journal}{Living Reviews in Relativity} \bibinfo{volume}{20},
  \bibinfo{pages}{3}.
\newblock \DOIprefix\doi{10.1007/s41114-017-0006-z},
  \href{http://arxiv.org/abs/1610.09381}{{\tt arXiv:1610.09381}}.
\bibitem[{{Metzger} and {Berger}(2012)}]{2012ApJ...746...48M}
\bibinfo{author}{{Metzger}, B.D.}, \bibinfo{author}{{Berger}, E.},
  \bibinfo{year}{2012}.
\newblock \bibinfo{title}{{What is the Most Promising Electromagnetic
  Counterpart of a Neutron Star Binary Merger?}}
\newblock \bibinfo{journal}{\apj} \bibinfo{volume}{746}, \bibinfo{pages}{48}.
\newblock \DOIprefix\doi{10.1088/0004-637X/746/1/48},
  \href{http://arxiv.org/abs/1108.6056}{{\tt arXiv:1108.6056}}.
\bibitem[{{Metzger} et~al.(2010){Metzger}, {Mart{\'{\i}}nez-Pinedo}, {Darbha}
  et~al.}]{2010MNRAS.406.2650M}
\bibinfo{author}{{Metzger}, B.D.}, \bibinfo{author}{{Mart{\'{\i}}nez-Pinedo},
  G.}, \bibinfo{author}{{Darbha}, S.}, et~al., \bibinfo{year}{2010}.
\newblock \bibinfo{title}{{Electromagnetic counterparts of compact object
  mergers powered by the radioactive decay of r-process nuclei}}.
\newblock \bibinfo{journal}{\mnras} \bibinfo{volume}{406},
  \bibinfo{pages}{2650--2662}.
\newblock \DOIprefix\doi{10.1111/j.1365-2966.2010.16864.x},
  \href{http://arxiv.org/abs/1001.5029}{{\tt arXiv:1001.5029}}.
\bibitem[{{Morgan} et~al.(2019){Morgan}, {Bechtol}, {Kessler}
  et~al.}]{2019ApJ...883..125M}
\bibinfo{author}{{Morgan}, R.}, \bibinfo{author}{{Bechtol}, K.},
  \bibinfo{author}{{Kessler}, R.}, et~al., \bibinfo{year}{2019}.
\newblock \bibinfo{title}{{A DECam Search for Explosive Optical Transients
  Associated with IceCube Neutrino Alerts}}.
\newblock \bibinfo{journal}{\apj} \bibinfo{volume}{883}, \bibinfo{pages}{125}.
\newblock \DOIprefix\doi{10.3847/1538-4357/ab3a45}.
\bibitem[{{Morganson} et~al.(2018){Morganson}, {Gruendl}, {Menanteau}
  et~al.}]{2018PASP..130g4501M}
\bibinfo{author}{{Morganson}, E.}, \bibinfo{author}{{Gruendl}, R.A.},
  \bibinfo{author}{{Menanteau}, F.}, et~al., \bibinfo{year}{2018}.
\newblock \bibinfo{title}{{The Dark Energy Survey Image Processing Pipeline}}.
\newblock \bibinfo{journal}{\pasp} \bibinfo{volume}{130},
  \bibinfo{pages}{074501}.
\newblock \DOIprefix\doi{10.1088/1538-3873/aab4ef},
  \href{http://arxiv.org/abs/1801.03177}{{\tt arXiv:1801.03177}}.
\bibitem[{{Nair} et~al.(2018){Nair}, {Bose} and {Saini}}]{2018PhRvD..98b3502N}
\bibinfo{author}{{Nair}, R.}, \bibinfo{author}{{Bose}, S.},
  \bibinfo{author}{{Saini}, T.D.}, \bibinfo{year}{2018}.
\newblock \bibinfo{title}{{Measuring the Hubble constant: Gravitational wave
  observations meet galaxy clustering}}.
\newblock \bibinfo{journal}{\prd} \bibinfo{volume}{98},
  \bibinfo{pages}{023502}.
\newblock \DOIprefix\doi{10.1103/PhysRevD.98.023502},
  \href{http://arxiv.org/abs/1804.06085}{{\tt arXiv:1804.06085}}.
\bibitem[{{Nissanke} et~al.(2013){Nissanke}, {Holz}, {Dalal}
  et~al.}]{2013arXiv1307.2638N}
\bibinfo{author}{{Nissanke}, S.}, \bibinfo{author}{{Holz}, D.E.},
  \bibinfo{author}{{Dalal}, N.}, et~al., \bibinfo{year}{2013}.
\newblock \bibinfo{title}{{Determining the Hubble constant from gravitational
  wave observations of merging compact binaries}}.
\newblock \bibinfo{journal}{arXiv e-prints} ,
  \bibinfo{pages}{arXiv:1307.2638}\href{http://arxiv.org/abs/1307.2638}{{\tt
  arXiv:1307.2638}}.
\bibitem[{{Nissanke} et~al.(2010){Nissanke}, {Holz}, {Hughes}
  et~al.}]{2010ApJ...725..496N}
\bibinfo{author}{{Nissanke}, S.}, \bibinfo{author}{{Holz}, D.E.},
  \bibinfo{author}{{Hughes}, S.A.}, et~al., \bibinfo{year}{2010}.
\newblock \bibinfo{title}{{Exploring Short Gamma-ray Bursts as
  Gravitational-wave Standard Sirens}}.
\newblock \bibinfo{journal}{\apj} \bibinfo{volume}{725},
  \bibinfo{pages}{496--514}.
\newblock \DOIprefix\doi{10.1088/0004-637X/725/1/496},
  \href{http://arxiv.org/abs/0904.1017}{{\tt arXiv:0904.1017}}.
\bibitem[{{Palmese} et~al.(2017){Palmese}, {Hartley}, {Tarsitano}
  et~al.}]{2017ApJ...849L..34P}
\bibinfo{author}{{Palmese}, A.}, \bibinfo{author}{{Hartley}, W.},
  \bibinfo{author}{{Tarsitano}, F.}, et~al., \bibinfo{year}{2017}.
\newblock \bibinfo{title}{{Evidence for Dynamically Driven Formation of the
  GW170817 Neutron Star Binary in NGC 4993}}.
\newblock \bibinfo{journal}{\apj} \bibinfo{volume}{849}, \bibinfo{pages}{L34}.
\newblock \DOIprefix\doi{10.3847/2041-8213/aa9660},
  \href{http://arxiv.org/abs/1710.06748}{{\tt arXiv:1710.06748}}.
\bibitem[{{Palmese} and {Kim}(2020)}]{2020arXiv200504325P}
\bibinfo{author}{{Palmese}, A.}, \bibinfo{author}{{Kim}, A.G.},
  \bibinfo{year}{2020}.
\newblock \bibinfo{title}{{Probing gravity and growth of structure with
  gravitational waves and galaxies' peculiar velocity}}.
\newblock \bibinfo{journal}{arXiv e-prints} ,
  \bibinfo{pages}{arXiv:2005.04325}\href{http://arxiv.org/abs/2005.04325}{{\tt
  arXiv:2005.04325}}.
\bibitem[{{Pan} et~al.(2017){Pan}, {Foley}, {Smith}
  et~al.}]{2017MNRAS.470.4241P}
\bibinfo{author}{{Pan}, Y.C.}, \bibinfo{author}{{Foley}, R.J.},
  \bibinfo{author}{{Smith}, M.}, et~al. (\bibinfo{collaboration}{The DES
  Collaboration}), \bibinfo{year}{2017}.
\newblock \bibinfo{title}{{DES15E2mlf: a spectroscopically confirmed
  superluminous supernova that exploded 3.5 Gyr after the big bang}}.
\newblock \bibinfo{journal}{\mnras} \bibinfo{volume}{470},
  \bibinfo{pages}{4241--4250}.
\newblock \DOIprefix\doi{10.1093/mnras/stx1467},
  \href{http://arxiv.org/abs/1707.06649}{{\tt arXiv:1707.06649}}.
\bibitem[{Paschalidis and Ruiz(2019)}]{PhysRevD.100.043001}
\bibinfo{author}{Paschalidis, V.}, \bibinfo{author}{Ruiz, M.},
  \bibinfo{year}{2019}.
\newblock \bibinfo{title}{Are fast radio bursts the most likely electromagnetic
  counterpart of neutron star mergers resulting in prompt collapse?}
\newblock \bibinfo{journal}{Phys. Rev. D} \bibinfo{volume}{100},
  \bibinfo{pages}{043001}.
\newblock \URLprefix
  \url{https://link.aps.org/doi/10.1103/PhysRevD.100.043001},
  \DOIprefix\doi{10.1103/PhysRevD.100.043001}.
\bibitem[{Pordes et~al.(2007)Pordes, Petravick, Kramer et~al.}]{Pordes_2007}
\bibinfo{author}{Pordes, R.}, \bibinfo{author}{Petravick, D.},
  \bibinfo{author}{Kramer, B.}, et~al., \bibinfo{year}{2007}.
\newblock \bibinfo{title}{The open science grid}.
\newblock \bibinfo{journal}{J. Phys.: Conf. Ser.} \bibinfo{volume}{78},
  \bibinfo{pages}{012057}.
\newblock \URLprefix
  \url{https://doi.org/10.1088%2F1742-6596%2F78%2F1%2F012057},
  \DOIprefix\doi{10.1088/1742-6596/78/1/012057}.
\bibitem[{{Pursiainen} et~al.(2018){Pursiainen}, {Childress}, {Smith}
  et~al.}]{2018MNRAS.481..894P}
\bibinfo{author}{{Pursiainen}, M.}, \bibinfo{author}{{Childress}, M.},
  \bibinfo{author}{{Smith}, M.}, et~al. (\bibinfo{collaboration}{The DES
  Collaboration}), \bibinfo{year}{2018}.
\newblock \bibinfo{title}{{Rapidly evolving transients in the Dark Energy
  Survey}}.
\newblock \bibinfo{journal}{\mnras} \bibinfo{volume}{481},
  \bibinfo{pages}{894--917}.
\newblock \DOIprefix\doi{10.1093/mnras/sty2309}.
\bibitem[{{Radice} et~al.(2018){Radice}, {Perego}, {Hotokezaka}
  et~al.}]{2018ApJ...869..130R}
\bibinfo{author}{{Radice}, D.}, \bibinfo{author}{{Perego}, A.},
  \bibinfo{author}{{Hotokezaka}, K.}, et~al., \bibinfo{year}{2018}.
\newblock \bibinfo{title}{{Binary Neutron Star Mergers: Mass Ejection,
  Electromagnetic Counterparts, and Nucleosynthesis}}.
\newblock \bibinfo{journal}{\apj} \bibinfo{volume}{869}, \bibinfo{pages}{130}.
\newblock \DOIprefix\doi{10.3847/1538-4357/aaf054},
  \href{http://arxiv.org/abs/1809.11161}{{\tt arXiv:1809.11161}}.
\bibitem[{{Riess} et~al.(2019){Riess}, {Casertano}, {Yuan}
  et~al.}]{2019ApJ...876...85R}
\bibinfo{author}{{Riess}, A.G.}, \bibinfo{author}{{Casertano}, S.},
  \bibinfo{author}{{Yuan}, W.}, et~al., \bibinfo{year}{2019}.
\newblock \bibinfo{title}{{Large Magellanic Cloud Cepheid Standards Provide a
  1\% Foundation for the Determination of the Hubble Constant and Stronger
  Evidence for Physics beyond {\ensuremath{\Lambda}}CDM}}.
\newblock \bibinfo{journal}{\apj} \bibinfo{volume}{876}, \bibinfo{pages}{85}.
\newblock \DOIprefix\doi{10.3847/1538-4357/ab1422},
  \href{http://arxiv.org/abs/1903.07603}{{\tt arXiv:1903.07603}}.
\bibitem[{{Rosswog} et~al.(2017){Rosswog}, {Feindt}, {Korobkin}
  et~al.}]{2017CQGra..34j4001R}
\bibinfo{author}{{Rosswog}, S.}, \bibinfo{author}{{Feindt}, U.},
  \bibinfo{author}{{Korobkin}, O.}, et~al., \bibinfo{year}{2017}.
\newblock \bibinfo{title}{{Detectability of compact binary merger macronovae}}.
\newblock \bibinfo{journal}{Class. and Quantum Gravity} \bibinfo{volume}{34},
  \bibinfo{pages}{104001}.
\newblock \DOIprefix\doi{10.1088/1361-6382/aa68a9},
  \href{http://arxiv.org/abs/1611.09822}{{\tt arXiv:1611.09822}}.
\bibitem[{{Schutz}(1986)}]{1986Natur.323..310S}
\bibinfo{author}{{Schutz}, B.F.}, \bibinfo{year}{1986}.
\newblock \bibinfo{title}{{Determining the Hubble constant from gravitational
  wave observations}}.
\newblock \bibinfo{journal}{Nature} \bibinfo{volume}{323},
  \bibinfo{pages}{310}.
\newblock \DOIprefix\doi{10.1038/323310a0}.
\bibitem[{Sfiligoi et~al.(2009)Sfiligoi, Bradley, Holzman et~al.}]{glideinwms}
\bibinfo{author}{Sfiligoi, I.}, \bibinfo{author}{Bradley, D.C.},
  \bibinfo{author}{Holzman, B.}, et~al., \bibinfo{year}{2009}.
\newblock \bibinfo{title}{The pilot way to grid resources using glideinwms},
  in: \bibinfo{booktitle}{Proceedings of the 2009 WRI World Congress on
  Computer Science and Information Engineering - Volume 02},
  \bibinfo{publisher}{IEEE Computer Society}, \bibinfo{address}{Washington, DC,
  USA}. pp. \bibinfo{pages}{428--432}.
\newblock \URLprefix \url{https://doi.org/10.1109/CSIE.2009.950},
  \DOIprefix\doi{10.1109/CSIE.2009.950}.
\bibitem[{{Singer} et~al.(2016){Singer}, {Chen}, {Holz} et~al.}]{Singer2016}
\bibinfo{author}{{Singer}, L.P.}, \bibinfo{author}{{Chen}, H.Y.},
  \bibinfo{author}{{Holz}, D.E.}, et~al., \bibinfo{year}{2016}.
\newblock \bibinfo{title}{{Supplement: {\ldquo}Going the Distance: Mapping Host
  Galaxies of LIGO and Virgo Sources in Three Dimensions Using Local
  Cosmography and Targeted Follow-up{\rdquo} (2016, ApJ, 829, L15)}}.
\newblock \bibinfo{journal}{Astrophys.~J.~s} \bibinfo{volume}{226},
  \bibinfo{pages}{10}.
\newblock \DOIprefix\doi{10.3847/0067-0049/226/1/10},
  \href{http://arxiv.org/abs/1605.04242}{{\tt arXiv:1605.04242}}.
\bibitem[{{Skrutskie} et~al.(2006){Skrutskie}, {Cutri}, {Stiening}
  et~al.}]{2006AJ....131.1163S}
\bibinfo{author}{{Skrutskie}, M.F.}, \bibinfo{author}{{Cutri}, R.M.},
  \bibinfo{author}{{Stiening}, R.}, et~al., \bibinfo{year}{2006}.
\newblock \bibinfo{title}{{The Two Micron All Sky Survey (2MASS)}}.
\newblock \bibinfo{journal}{\aj} \bibinfo{volume}{131},
  \bibinfo{pages}{1163--1183}.
\newblock \DOIprefix\doi{10.1086/498708}.
\bibitem[{{Soares-Santos} et~al.(2017){Soares-Santos}, {Holz}, {Annis}
  et~al.}]{2017ApJ...848L..16S}
\bibinfo{author}{{Soares-Santos}, M.}, \bibinfo{author}{{Holz}, D.E.},
  \bibinfo{author}{{Annis}, J.}, et~al., \bibinfo{year}{2017}.
\newblock \bibinfo{title}{{The Electromagnetic Counterpart of the Binary
  Neutron Star Merger LIGO/Virgo GW170817. I. Discovery of the Optical
  Counterpart Using the Dark Energy Camera}}.
\newblock \bibinfo{journal}{\apj} \bibinfo{volume}{848}, \bibinfo{pages}{L16}.
\newblock \DOIprefix\doi{10.3847/2041-8213/aa9059},
  \href{http://arxiv.org/abs/1710.05459}{{\tt arXiv:1710.05459}}.
\bibitem[{{Soares-Santos} et~al.(2016){Soares-Santos}, {Kessler}, {Berger}
  et~al.}]{2016ApJ...823L..33S}
\bibinfo{author}{{Soares-Santos}, M.}, \bibinfo{author}{{Kessler}, R.},
  \bibinfo{author}{{Berger}, E.}, et~al., \bibinfo{year}{2016}.
\newblock \bibinfo{title}{{A Dark Energy Camera Search for an Optical
  Counterpart to the First Advanced LIGO Gravitational Wave Event GW150914}}.
\newblock \bibinfo{journal}{\apj} \bibinfo{volume}{823}, \bibinfo{pages}{L33}.
\newblock \DOIprefix\doi{10.3847/2041-8205/823/2/L33},
  \href{http://arxiv.org/abs/1602.04198}{{\tt arXiv:1602.04198}}.
\bibitem[{{Soares-Santos \& Palmese} et~al.(2019){Soares-Santos \& Palmese},
  Hartley et~al.}]{Soares_Santos_2019}
\bibinfo{author}{{Soares-Santos \& Palmese}}, \bibinfo{author}{Hartley, W.},
  et~al., \bibinfo{year}{2019}.
\newblock \bibinfo{title}{First measurement of the hubble constant from a dark
  standard siren using the dark energy survey galaxies and the {LIGO}/virgo
  binary{\textendash}black-hole merger {GW}170814}.
\newblock \bibinfo{journal}{\apj} \bibinfo{volume}{876}, \bibinfo{pages}{L7}.
\newblock \URLprefix \url{https://doi.org/10.3847%2F2041-8213%2Fab14f1},
  \DOIprefix\doi{10.3847/2041-8213/ab14f1}.
\bibitem[{{Tanaka} and {Hotokezaka}(2013)}]{2013ApJ...775..113T}
\bibinfo{author}{{Tanaka}, M.}, \bibinfo{author}{{Hotokezaka}, K.},
  \bibinfo{year}{2013}.
\newblock \bibinfo{title}{{Radiative Transfer Simulations of Neutron Star
  Merger Ejecta}}.
\newblock \bibinfo{journal}{\apj} \bibinfo{volume}{775}, \bibinfo{pages}{113}.
\newblock \DOIprefix\doi{10.1088/0004-637X/775/2/113},
  \href{http://arxiv.org/abs/1306.3742}{{\tt arXiv:1306.3742}}.
\bibitem[{{Tanaka} et~al.(2018){Tanaka}, {Kato}, {Gaigalas}
  et~al.}]{2018ApJ...852..109T}
\bibinfo{author}{{Tanaka}, M.}, \bibinfo{author}{{Kato}, D.},
  \bibinfo{author}{{Gaigalas}, G.}, et~al., \bibinfo{year}{2018}.
\newblock \bibinfo{title}{{Properties of Kilonovae from Dynamical and
  Post-merger Ejecta of Neutron Star Mergers}}.
\newblock \bibinfo{journal}{\apj} \bibinfo{volume}{852}, \bibinfo{pages}{109}.
\newblock \DOIprefix\doi{10.3847/1538-4357/aaa0cb},
  \href{http://arxiv.org/abs/1708.09101}{{\tt arXiv:1708.09101}}.
\bibitem[{{Villar} et~al.(2017){Villar}, {Guillochon}, {Berger}
  et~al.}]{2017ApJ...851L..21V}
\bibinfo{author}{{Villar}, V.A.}, \bibinfo{author}{{Guillochon}, J.},
  \bibinfo{author}{{Berger}, E.}, et~al., \bibinfo{year}{2017}.
\newblock \bibinfo{title}{{The Combined Ultraviolet, Optical, and Near-infrared
  Light Curves of the Kilonova Associated with the Binary Neutron Star Merger
  GW170817: Unified Data Set, Analytic Models, and Physical Implications}}.
\newblock \bibinfo{journal}{\apjl} \bibinfo{volume}{851}, \bibinfo{pages}{L21}.
\newblock \DOIprefix\doi{10.3847/2041-8213/aa9c84},
  \href{http://arxiv.org/abs/1710.11576}{{\tt arXiv:1710.11576}}.
\bibitem[{{Young}(1994)}]{1994ApOpt..33.1108Y}
\bibinfo{author}{{Young}, A.T.}, \bibinfo{year}{1994}.
\newblock \bibinfo{title}{{Air mass and refraction}}.
\newblock \bibinfo{journal}{Appl. Opt.} \bibinfo{volume}{33},
  \bibinfo{pages}{1108--1110}.
\newblock \DOIprefix\doi{10.1364/AO.33.001108}.
\bibitem[{{Zacharias} et~al.(2004){Zacharias}, {Monet}, {Levine}
  et~al.}]{2004AAS...205.4815Z}
\bibinfo{author}{{Zacharias}, N.}, \bibinfo{author}{{Monet}, D.G.},
  \bibinfo{author}{{Levine}, S.E.}, et~al., \bibinfo{year}{2004}.
\newblock \bibinfo{title}{{The Naval Observatory Merged Astrometric Dataset
  (NOMAD)}}, in: \bibinfo{booktitle}{American Astronomical Society Meeting
  Abstracts}, p. \bibinfo{pages}{48.15}.

\end{thebibliography}
\bibliographystyle{elsarticle-harv}
\end{document}